\documentclass[epj,final]{svjour}
\usepackage{graphicx}
\usepackage{amsmath}
\def\be{\begin{eqnarray}}
\def\ee{\end{eqnarray}}
\def\rb{{\bar r}}

\begin{document}
%=================================================================%
\title{Higher-order anisotropies in the Blast-Wave Model -- disentangling flow and density field anisotropies}
\author{Jakub Cimerman\inst{1,2}, Boris Tom\'a\v{s}ik\inst{1,3}, M\'at\'e Csan\'ad\inst{4}, and S\'andor L\"ok\"os\inst{4}}
\institute{
Czech Technical University in Prague, FNSPE, B\v{r}ehov\'a 7,
11519 Prague 1, Czech Republic
\and
Comenius University, FMPI, Mlynsk\'a Dolina F1, 84248 Bratislava, Slovakia
\and
Univerzita Mateja Bela, FPV, Tajovsk\'eho 40,
97401 Bansk\'a Bystrica, Slovakia
 \and E\"otv\"os Lor\'and University, P\'azm\'any P. s. 1/a, H-1117 Budapest,  Hungary
}
\titlerunning{Anisotropies in the Blast-Wave Model}
\authorrunning{J. Cimerman, B. Tom\'a\v{s}ik, M. Csan\'ad, and S. L\"ok\"os} 
%=================================================================%
\date{}
\abstract{
We formulate a generalisation of the blast-wave model which is suitable for the description
of higher order azimuthal anisotropies of the hadron production. The model includes anisotropy 
in the density profile as well as an anisotropy in the transverse expansion velocity field. We then 
study how these two kinds of anisotropies influence the single-particle distributions and the 
correlation radii of two-particle correlation functions. Particularly we focus on the third-order 
anisotropy and consideration is given  averaging over different orientations of the event plane. 
}

\maketitle
%=======================================================================

\section{Introduction}

The hot matter excited in ultrarelativistic heavy-ion collisions at colliders like the LHC or RHIC exhibits
a sizeable anisotropy in particle production perpendicularly to the beam direction
\cite{Abelev:2014pua,ATLAS:2012at,Chatrchyan:2012ta,Abelev:2008ae,Adler:2003kt}. 
The azimuthal anisotropy of
hadron momentum distributions, measured in terms of Fourier coefficients, is caused by the anisotropy of 
the fireball at freeze-out in spatial density and expansion pattern. That, in turn,  results from its evolution which 
starts from an\-iso\-tro\-pic initial conditions 
\cite{Gale:2012rq,Niemi:2012aj,Floerchinger:2013hza,Qian:2013nba,Heinz:2013wva}
and may further receive anisotropic excitations on the way
\cite{Schulc:2014jma,Nahrgang:2011vn,Tachibana:2017syd,Crkovska:2016flo}. The 
evolution depends on the Equation of State and the transport coefficients
\cite{Luzum:2009sb,Ryu:2015vwa,Denicol:2015nhu}. Thus by measuring the 
final state anisotropies one gets an access to the intrinsic properties of the matter
\cite{Bernhard:2016tnd}. Note that the anisotropies are 
% BT: Mate wanted to add which kinds of anisotropies we mean, but I think that the sentence would become too cumbersome
unique in each event and a large fraction of them, especially higher-order anisotropies, 
are averaged out if measured in a  sample consisting of a large 
number of events. 

% BT: I dropped Mate's comment about spontaneous emission being approximation. I think it does not 
% contribute to the clarity of the text.
Hadrons are emitted at the moment of freeze-out and this is when their distributions 
are formed. Hence, the direct information that they are carrying is about the state of the fireball at this very 
moment. The present paper deals with this relation. It is then the task for hydrodynamic and/or transport 
simulations to conclude about the preceding evolution of the fireball. 

There are two kinds of anisotropies of the fireball at the moment of freeze-out, that may cause an
anisotropy of the hadron distribution
\cite{Tomasik:2004bn,Csanad:2008af,Plumberg:2013nga,Lokos:2016fze}. 
Firstly, if the transverse expansion velocity in some directions is 
higher than in the others, the stronger blueshift of the momentum spectra in those directions will cause 
momentum anisotropy. Secondly, alone an expanding and spatially anisotropic fireball may also produce an 
anisotropy of the momentum distribution. 
Unfortunately, the two mechanisms cannot be distinguished by mere measurement of the momentum
anisotropies. 

For the second-order anisotropies it has been shown that the solution is offered by the azimuthal dependence 
of the femtoscopic correlation radii, which are more sensitive to spatial anisotropy. Detailed studies with the
help of blast-wave \cite{Tomasik:2004bn} and Buda-Lund  \cite{Csanad:2008af} models have been performed. 
After PHENIX 
has published the azimuthal dependence of the correlation radii \cite{Niida:2013lia} with respect to the 
\emph{third-order} event plane, 
the problem has been reconsidered at that order in \cite{Plumberg:2013nga}. It has been demonstrated 
in framework of a toy model that at RHIC the 
spatial anisotropy is the driving feature which determines the phase of the oscillation of the correlation radii.
This is in agreement with the second-order results \cite{Tomasik:2004bn,Csanad:2008af}.
Today, higher experimental statistics allows for more detailed investigations and the data 
on third-order azimuthal dependence of correlation radii call for more detailed theoretical studies
\cite{Adams:2003ra,Adare:2014vax,Loggins:2014uaa}.
The third-order 
anisotropy together with azimuthal dependence of the correlation radii have been investigated by some 
of us 
in detail in framework of the Buda-Lund model recently \cite{Lokos:2016fze}. The dependence of the 
oscillation amplitudes on parameters of the model which gauge the anisotropies in space and flow has
been calculated in great detail up to 6th order.

The present paper is analogical to \cite{Lokos:2016fze}. We reconsider the problem in framework of the 
blast-wave model. This is perhaps the most commonly used model for the 
analysis of soft hadron data in high energy heavy-ion physics. Therefore, we
first systematically extend it for anisotropies  of higher order, although we later use it only up to the third 
order. By doing this we actually modify the formulation of the model as proposed in \cite{Retiere:2003kf},
since that one was adequate only up to second order. Note that our extension also 
slightly differs from the one proposed in \cite{Adare:2014kci}, since want to allow for  varying 
transverse size of the fireball.

Then, with the generalised blast-wave model we investigate in quite detail how the oscillation amplitudes of 
the correlation radii depend on parameters which measure the flow and the shape anisotropy. On top of the 
schematic picture obtained in \cite{Plumberg:2013nga} we add the details by providing parameter maps, i.e.\ 
contour plots of the dependence of oscillation amplitudes on both flow and space anisotropy, similarly 
as was done for the Buda-Lund model in \cite{Lokos:2016fze}. Such maps should allow, at least in principle, 
to infer both  values for flow and space anisotropy from the measured data.
We also go beyond the toy-model study 
of \cite{Plumberg:2013nga} by including both the second and the third order anisotropies into the calculation
and integrating over the order not actually being investigated, analogically to real experiment.

In the next Section we introduce and explain the extension of the blast-wave model used in this study. 
Then, Section~\ref{s:vn} is devoted to calculations of the anisotropies of single-particle distributions. 
Oscillations of correlation radii are investigated in 
Section~\ref{s:femto}. 
We demonstrate our results with the help of qualitative data analysis  in Section \ref{s:data}.
All results are summarised in the concluding Section. Some technical 
details are explained in the Appendices.

%%%%%%%%%%%%%%%%%%%%%%%%%%%%%%%%%%%%%%%%%%%%%%%%%%

\section{The extended blast-wave model}
\label{s:bw}

Particle production is described with the help of the emission function $S(x,p)$, which is 
the Wigner function, i.e.\ the phase-space density 
of hadrons that are being emitted from the fireball. In the blast-wave model 
\cite{Siemens:1978pb,Schnedermann:1993ws,Csorgo:1995bi,Tomasik:1999cq,Retiere:2003kf}
it is parametrised as
\begin{multline}
S(x,p) d^4x = \frac{g}{(2\pi)^3} \, 
m_t\cosh(Y - \eta_s)
\\
 r\, dr\, d\theta\, \tau \, d\eta \, \frac{d\tau}{\sqrt{2\pi}\Delta\tau}
\exp\left ( - \frac{(\tau - \tau_0)^2}{2\Delta\tau^2}  \right )\\
\Theta(r - R(\theta)) \,
\exp\left (- \frac{p^\mu u_\mu}{T}  \right )\,  .
\end{multline}
Here, $g$ is the spin degeneracy factor, $T$ is the local temperature, 
$u^\mu$ is the expansion velocity field,
and $\Theta(x)$ is the Heaviside step function.
We parametrise the momentum of a particle 
with the help of transverse momentum $p_t$, transverse mass $m_t$, rapidity $Y$
and the azimuthal angle $\phi$ as
\begin{equation}
p^\mu = ( m_t \cosh Y, p_t \cos \phi, p_t \sin\phi, m_t \sinh Y)\,  . 
\end{equation}
In this paper we shall denote the rapidity with capital letter in order to distinguish it 
from the spatial coordinate. 

As spatial coordinates we use the radial coordinate $r$ and the azimuthal angle $\theta$, 
%BT: I'll keep in theta for the azimuthal angle. alpha is even less frequently used. 
as well as the space-time rapidity $\eta_s = \frac{1}{2}\ln \left ( (t+z)/(t-z)\right )$ and longitudinal 
proper time $\tau =\sqrt{t^2 - z^2}$. Then
\begin{equation}
x^\mu = (\tau \cosh \eta_s, r \cos\theta, r\sin\theta, \tau \sinh\eta_s)\,  .
\end{equation}
% BT: equations are numbered only if we may need to refer to it in further text

Furthermore, $R(\theta)$ is the transverse size of the fireball, 
depending on the azimuthal angle $\theta$.
The spatial anisotropy of the model is specified by the particular prescription for $R(\theta)$. 
The transverse size is then parametrised as
\begin{equation}
\label{e:Rtheta}
R(\theta) = R_0 \left ( 1 - \sum_{n=2}^\infty a_n \cos(n(\theta - \theta_n))   \right )\,  ,
\end{equation}
where the amplitudes $a_n$ and the phases $\theta_n$ are model parameters. 
Note that $\theta_n$'s denote the orientations of the so-called $n$-th order event planes. Note that 
in the series we have skipped the first-order term which leads to mere shift of the shape. 
Also note that the amplitudes for the oscillation are pa\-ra\-me\-tri\-sed in an unusual way, with the 
help of $(-a_n)$. In this way, the resulting $v_n$ is to first order proportional to $a_n$, as 
will be seen later.

Note that in \cite{Tomasik:2004bn} a different parametrisation for the elliptic shape 
of the fireball was used, with radii $R_x$ and $R_y$ along the two axes of the ellipse
\begin{equation}
\label{e:aparam}
R_x = a R'\, , \qquad R_y = \frac{R'}{a}\,  ,
\end{equation}
with $R'$ and $a$ being parameters of the model. The advantage of that parametrisation 
is that when $a$ is being tuned, the volume stays constant and proportional to $R^2$. 
Nevertheless, such a prescription cannot be naturally generalised to higher orders. Therefore, 
we will now use Eq.~(\ref{e:Rtheta}). Some comments on the relation of the two parametrisations
can be found in Appendix~\ref{a:a1}.

It is convenient for further calculation to define a dimensionless transverse coordinate
\begin{equation}
\rb = \frac{r}{R(\theta)}\,  .
\label{e:rb}
\end{equation}
Particle production in our model occurs for $\rb$ in the range $[0,1]$.
%which assumes the values $0\le \rb \le 1$.

Expansion is described by the velocity field $u^\mu$. Velocity includes  longitudinal as well as
transverse component
\begin{multline}
u^\mu = (\cosh\eta_s\cosh\rho, \sinh\rho \cos\theta_b, 
\\
\sinh\rho \sin\theta_b, \sinh\eta_s \cosh\rho)
\end{multline}
where 
\begin{equation}
\theta_b = \theta_b(r,\theta)
\end{equation}
% BT: in this case it is more practical to use r and not rbar
is the angle of the transverse vector of the velocity and will be specified below. Furthermore 
\begin{equation}
\rho = \rho(\rb, \theta_b)
\end{equation}
is the rapidity connected with the transverse velocity, so that the transverse velocity at 
midrapidity is $v_t = \tanh\rho$.

The canonical (azimuthally symmetric) blast-wave mo\-del  is recovered if 
$R(\theta)$ and $\rho(\rb,\theta)$ do not depend on the angle and $\theta_b = \theta$. 
Here we construct the extension to arbitrary order of anisotropy. 

In a fireball without azimuthal symmetry we must specify the direction of the transverse
expansion velocity. In \cite{Tomasik:2004bn}, two models were investigated which differed
in the choice of that direction. Note, however, that only second-order anisotropy was studied 
there. It turned out that femtoscopic data \cite{Adams:2003ra} agreed with the choice 
in which the transverse velocity was always perpendicular to the surface of the fireball. 
We adopt this choice also here. Transverse velocity will be locally perpendicular to 
surfaces with constant $\rb$. Note that this is the natural direction of transverse pressure gradient 
and thus the acceleration. Hence, we actually identify the direction of the velocity with that of
acceleration. Such a choice is expected to be valid if the fireball decouples fast. 

The azimuthal angle of the velocity $\theta_b$ is then obtained from 
\begin{equation}
\label{e:thetab}
\tan \left ( \theta_b - \frac{\pi}{2}\right )
= \frac{dy}{dx} = \frac{\frac{dy}{d\theta}}{\frac{dx}{d\theta}}\,  ,
\end{equation}
where the derivative is taken along a surface with constant $\rb$. The solution is straightforward
and we summarise it together with the final result for $\theta_b$ in Appendix~\ref{a:a2}.

Finally, let us define the magnitude of the transverse velocity, which is parametrised with the help 
of transverse rapidity
\begin{equation}
\rho(\rb,\theta_b) = \rb \rho_0 \left (
1 + \sum_{n=2}^{\infty} 2 \rho_n \cos \left ( n (\theta_b - \theta_n)\right )
\right )\,  .
\end{equation}
Overall transverse flow is tuned with the help of $\rho_0$ and the anisotropies have 
amplitudes $\rho_n$. Note that we choose the same phase factors $\theta_n$ as we did for 
the spatial anisotropy. They are related to the event planes measured experimentally.
Note also the introduction of the factor 2 before $\rho_n$, unlike in Eq.~(\ref{e:Rtheta}).

In this paper we will restrict ourselves to anisotropies up to third order; higher orders will 
be omitted. 

Note also that we did not include corrections to the momentum distribution due to viscosity \cite{Teaney:2003kp}, 
as done e.g.~in \cite{Jaiswal:2015saa,Yang:2016rnw}. We plan to investigate this important issue 
in the future. 

%%%%%%%%%%%%%%%%%%%%%%%%%%%%%%%%%%%%%%%%%%%%%%%%%%

\section{Anisotropy of single-particle distributions}
\label{s:vn}

The single-particle spectrum is obtained by integrating the emission function
\begin{equation}
N_1(p_t,\phi,Y) = \frac{d^3N}{p_t\, dp_t\, dY\, d\phi} = \int S(x,p) d^4x\, .
\label{e:N1}
\end{equation}
The normalisation is such that the integral of $N_1(p_t,\phi,Y)$ over all momenta gives the number 
of particles.  
For the integrations of transverse directions in eq.~(\ref{e:N1}) 
it would be convenient to use polar coordinates
$r$ and $\theta$. However, it is even more convenient  to use $\rb$, defined 
in Eq.~(\ref{e:rb}), instead of $r$.
This requires a new Jacobian
\begin{equation}
r\, dr\, d\theta = \rb\, R^2(\theta)\, d\rb\, d\theta\,  .
\end{equation}

\begin{figure}
\begin{center}
\includegraphics[width=0.48\textwidth]{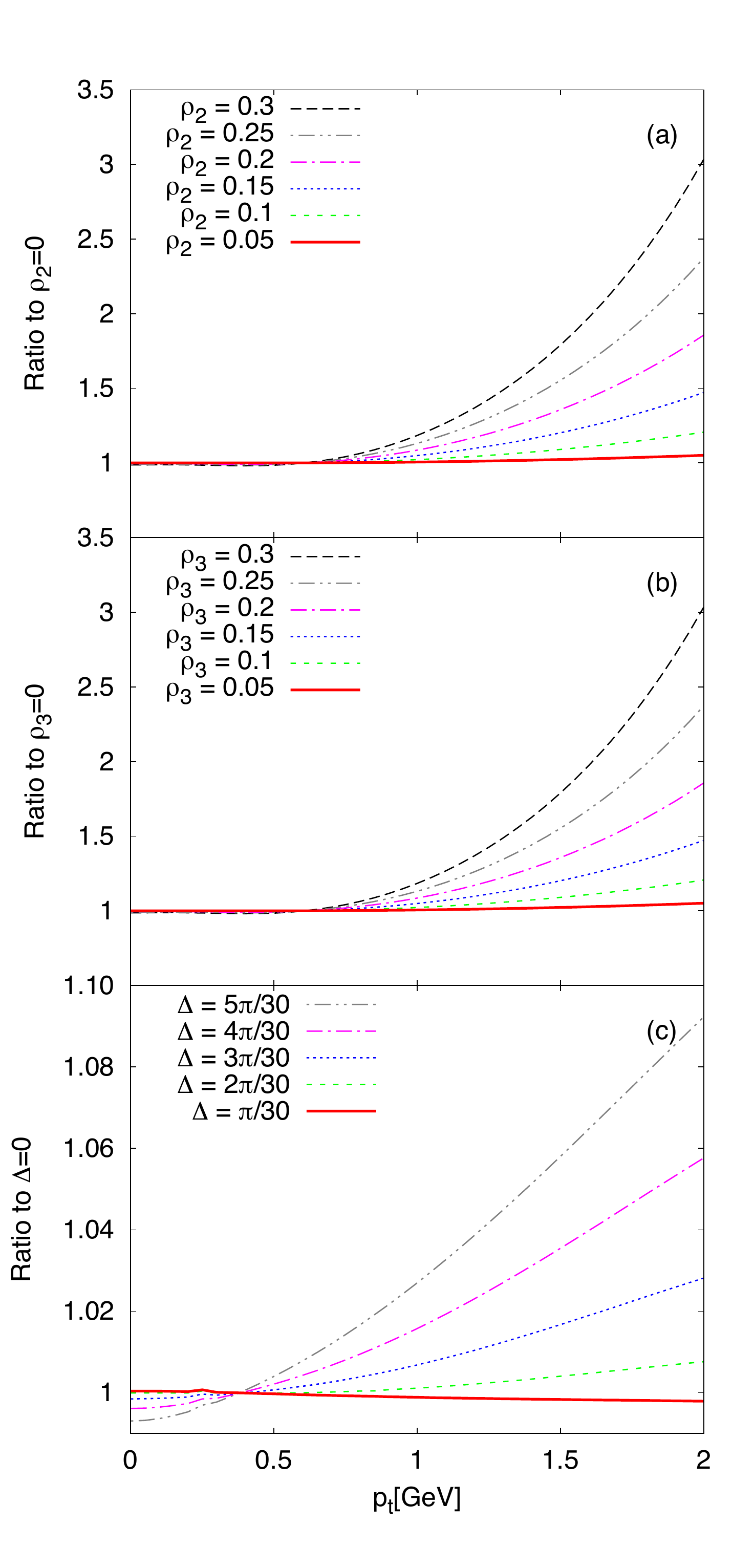}
\end{center}
\caption{The ratios of azimuthally integrated single particle spectra from fireballs with anisotropies
to a reference spectrum calculated for the same set parameters except all anisotropy 
coefficients set to 0. Calculated for (directly produced) pions and $T=120$~MeV, $\rho_0 = 0.8$,
$R_0 = 7$~fm, $\tau_0 = 10$ fm/$c$. (a) Ratios of spectra with second-order flow anisotropy 
and $a_2=a_3=\rho_3=0$. (b) Ratios of spectra with third-order flow anisotropy and 
$a_2=a_3=\rho_2=0$. (c) Ratios of spectra with $a_2=a_3=\rho_2=\rho_3=0.1$ and 
different values of $\Delta = \theta_3 - \theta_2$.}
\label{f:integspec}
\end{figure}

Before moving on towards the anisotropy it is interesting to explore if and how introducing spatial 
and flow anisotropy into the model modifies the \emph{azimuthally integrated} 
single-particle spectrum. We have checked that if we introduce only a spatial anisotropy 
(i.e.\ the $a_n$ coefficients may be non-vanishing, but all $\rho_n$'s are set to 0), 
then the normalisation may be slightly modified but the slope is unchanged.
%purely spatial anisotropy in this model 
%(i.e.\ only due to non-vanishing coefficients $a_n$ and all $\rho_n$'s set to 0) only 
%leads to a small change of the normalisation but no change in the slope. 
This is not the case 
for the flow anisotropy, however. 
As shown in Fig.~\ref{f:integspec}a and Fig.~\ref{f:integspec}b, flow anisotropy leads to slightly flatter spectra. 
%
%%%%%%%%%%%%%%%%%%%%%%
%%%%%%%%%%%%%%%%%%%%%%
%
We also show in Fig.~\ref{f:integspec}c
how the azimuthally integrated spectrum depends on the phase difference
of the second and third order event planes (cf. Eq.~(\ref{e:Rtheta}))
\begin{equation}
\Delta = \theta_3-\theta_2\,   .
\end{equation}
We do not expect any correlation  between the second-order and the third-order event planes
and to our knowledge there is no such correlation seen in the data.
Hence, all phase differences are realised equally likely. In a data sample averaged over a large number 
of events the mean value of all observed curves would be measured. 

Now we move on to the anisotropies of spectra which will be obtained as 
\begin{equation}
v_n = \frac{
\int N_1 (p_t,\phi,Y)\, \cos(n(\phi-\theta_n))\, d\phi
}{
\int N_1 (p_t,\phi,Y)\,  d\phi
}\,  .
\end{equation}
The single-particle distributions are calculated via Eq.~(\ref{e:N1}).
The anisotropy coefficients $v_n$ can be then expressed as 
\begin{equation}
v_n(p_t) = \frac{C_n(p_t)}{C_0(p_t)}
\end{equation}
where 
\begin{multline}
C_n(p_t) = \int_0^1 d\rb \int_0^{2\pi}d\theta\, \rb \, R^2(\theta) \cos(n(\theta_b(\theta)-\theta_n))
\\ \times
I_n\left ( \frac{p_t \sinh\rho(\rb,\theta)}{T} \right )
K_1\left ( \frac{p_t \cosh\rho(\rb,\theta)}{T} \right )
\end{multline}
where $I_n$ and $K_1$ are modified Bessel functions
and the integration over the azimuthal angle of the momentum $\phi$ was already performed here.

The result of the calculation, however, would depend on the value of the phase difference $\Delta$.
This dependence is hidden in $R(\theta)$ and $\rho(\rb,\theta)$. 
In an experimental analysis, one effectively takes an average over all its possible values. 
It has been shown in \cite{Lokos:2016fze} that the averaging may have an effect on the results.
We thus have to add this averaging and introduce
\begin{multline}
\bar C_n(p_t) = \int_0^{\frac{2}{3}\pi} d\Delta 
\int_0^1 d\rb \int_0^{2\pi}d\theta\, \rb\, R^2(\theta)
\\ \times \cos(n(\theta_b(\theta)-\theta_n))
\\ \times
I_n\left ( \frac{p_t \sinh\rho(\rb,\theta)}{T} \right )
K_1\left ( \frac{p_t \cosh\rho(\rb,\theta)}{T} \right )\,  .
\end{multline}
Then, the event-averaged $v_n$ is obtained as 
\begin{equation}
v_n(p_t) = \frac{\bar C_n(p_t)}{\bar C_0(p_t)}\,  .
\end{equation}
%This formula has been used in obtaining the results. 

\begin{figure}
\begin{center}
\includegraphics[width=0.48\textwidth]{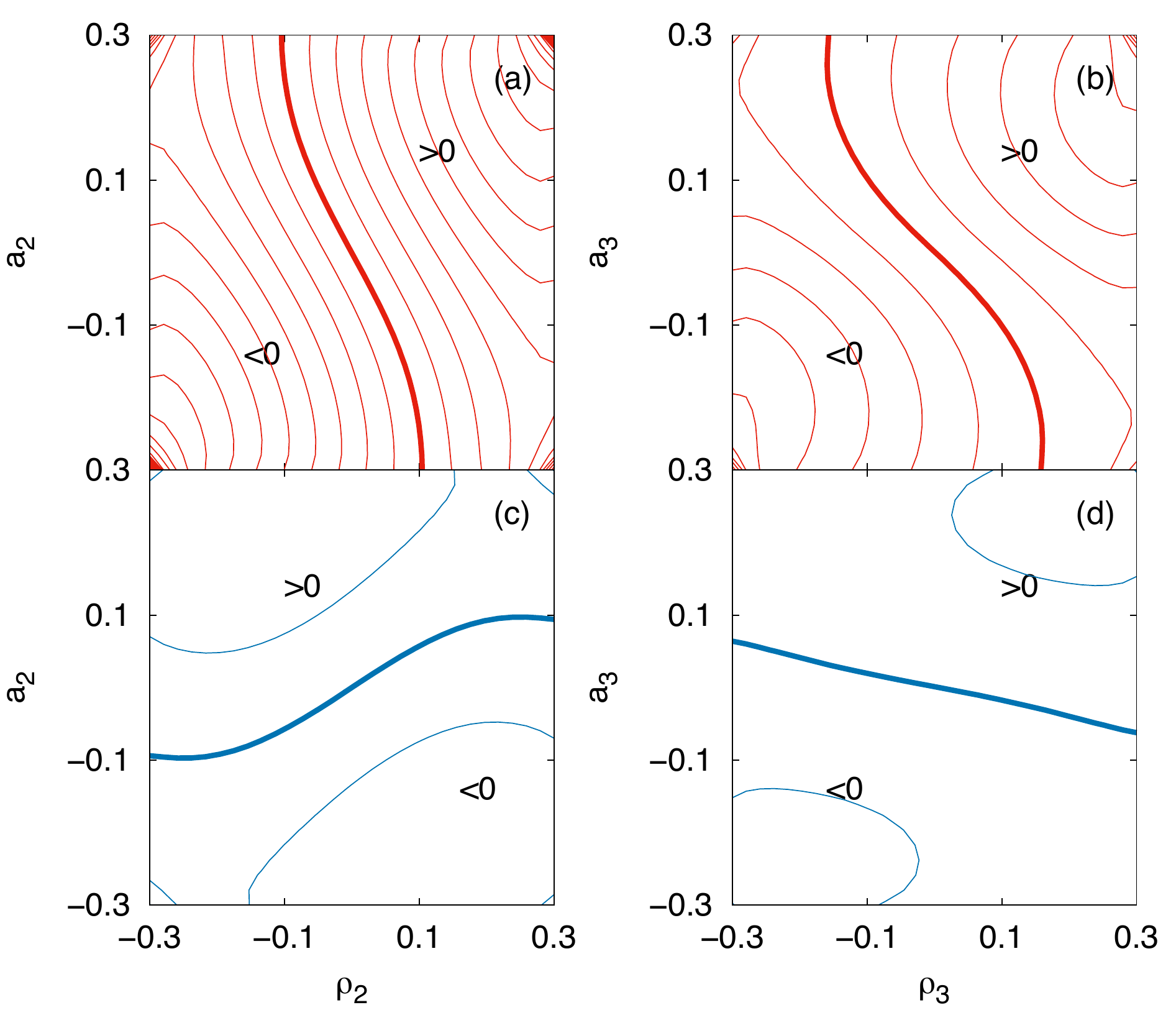}
\end{center}
\caption{
Dependence of $v_n$'s on spatial anisotropy $a_n$ and flow anisotropy $\rho_n$
of the same order. Results are shown for pions (upper row) and protons (lower row)
at $p_t = 300$~MeV. Model parameters are $T=120$~MeV, $\rho_0 = 0.8$,
$R_0 = 7$~fm, $\tau_0 = 10$ fm/$c$. The thick line identify $v_2$ or $v_3$ equal to 0, 
the thin lines  correspond to increment or decrease of $v_2$ ($v_3$) by 0.005.
}
\label{f:v2cntr}
\end{figure}

We calculated the dependence of $v_2$ and $v_3$ on the anisotropy coefficients $a_n$ and 
$\rho_n$. We have checked that the $v_n$'s of given order basically depend only on coefficients of the 
same order, therefore we shall only investigate such same-order dependences. 
In Fig.~\ref{f:v2cntr} we show the contour plots where dependence on both 
spatial and flow anisotropy can be seen. 
%
%%%%%%%%%%%%%%%%%%%%%%
%%%%%%%%%%%%%%%%%%%%%%
%
A complex structure is observed. In general, we can conclude, similarly to 
\cite{Tomasik:2004bn,Csanad:2008af,Lokos:2016fze}, that alone by measuring the anisotropy 
of single-particle distribution one is unable to determine uniquely both spatial and flow anisotropy. 
For heavy particles, like protons, $v_n$'s seem to be driven by the spatial anisotropy. Nevertheless, 
flow anisotropy kicks in as the value of $a_n$ grows. There is even a maximum assumed by 
$v_2$ and $v_3$ as a function of the corresponding $a_n$, although it may well be beyond the 
phenomenologically relevant parameter region. The reason is that for high enough  values 
of $a_n$ the outer surface of the fireball becomes concave and a smaller region moves 
transversely in the direction of the event plane. Although a unique combination of source parameters
might possibly be determined from the combination of measurements with different particle 
species, a clear 
answer shall be provided by measuring the azimuthal dependence of the correlation radii.

%%%%%%%%%%%%%%%%%%%%%%%%%%%%%%%%%%%%%%%%%%%%%%%%%%

\section{Spatial anisotropy seen by femtoscopy}
\label{s:femto}

The femtoscopic technique which uses two-particle correlations is standard tool for 
measuring the space-time characteristics of the emitting source. 
Here we employ the standard formalism where the correlation function is defined 
as 
\begin{equation}
C(p_1,p_2) = \frac{N_2(p_1,p_2)}{N_1(p_1) N_1(p_2)}
= \frac{\frac{d^6N}{dp_1^3 dp_2^3}}%
{\frac{d^3N}{dp_1^3}\frac{d^3N}{dp_2^3}}\,  .
\end{equation}
Instead of the momenta of the two particles, the correlation function is usually parametrised 
in terms of the average momentum $K$ and the momentum difference $q$
\begin{subequations}
\begin{eqnarray}
K & = & \frac{1}{2} (p_1 + p_2)\\
q & = & p_1 - p_2\,  .
\end{eqnarray}
\end{subequations}
Due to the on-shell constraint $K\cdot q = 0$, the time component $q^0$ can be expressed as
\begin{equation}
q^0 = \frac{\vec K}{K^0} \vec q = \vec \beta \vec q\,  .
\end{equation}
Hence, only three spatial components of $q$ shall be 
taken as independent. Note that we have introduced the pair velocity $\vec\beta$. 
The analysis is performed in the standard 
\emph{out-side-longitudinal reference frame} where the outward direction is identified with the 
direction of the transverse component of $\vec K$ and the longitudinal axis is parallel to the beam.
%\cite{r:OSLsystem}. 
Correlation function is then measured for $\vec K$ from some interval
and its inverse widths in $q$ carry 
information about space-time structure of the source. At given $\vec K$-range one does 
not measure the size of the whole fireball but rather its homogeneity lengths. Those are
the sizes of  homogeneity regions. The homogeneity region is a part 
of the whole fireball which produces hadrons 
with momentum $K$ from a given range. Due to expansion it is usually smaller than the whole fireball. 
We shall particularly look at how the homogeneity lengths vary with the azimuthal angle 
$\phi$ of the $\vec K$ vector. 

%Because in reality the shapes of the fireballs change from one event to another, events must be aligned 
%with respect to an event plane of  some order. Otherwise event averaging leads to azimuthally 
%symmetric effective source of particles.  If second-order event plane is used, the homogeneity 
%lengths measured over a large sample of events will exhibit second-order oscillation with the azimuthal angle with 
%possible admixture of higher even orders.  
%With alignment according to third-order event plane the oscillation will be of third (and higher multiples of 3) 
%order. The model formulated in this work describes such an effective source which is averaged over a large
%sample of aligned events.  

We will assume in what follows that the dependence of the correlation function on momentum 
difference $q$ can  be reasonably well parametrised by a Gaussian prescription 
\begin{multline}
C(q,K) -1 = \exp\bigl (
-R_o^2 q_o^2 - R_s^2 q_s^2 - R_l^2q_l^2 
\\
- 
2R_{os}^2 q_o q_s - 2R_{ol}^2 q_o q_l - 2 R_{sl}^2 q_s q_l
\bigr )\,  ,
\end{multline}
where $R_o$, $R_s$, $R_l$,  $R_{os}$, $R_{ol}$, and $R_{sl}$ are the correlation radii which can 
depend on $K$. They will be directly calculated from the emission function, see below.
  
It is important to realise that the coordinate frame in which correlations are measured is specified 
by the hadrons used in the measurement. It is different from any coordinate system 
which is attached to the fireball. 
The rotation between the two frames
defines the \emph{explicit} angular dependence of the correlation radii. In addition to this, due to 
collective expansion of the fireball, hadrons flying in different directions come from different 
parts of the fireball and carry information about their  homogeneity lengths. This 
introduces the \emph{implicit} angular dependence of the correlation radii \cite{Heinz:2002au}.

Generally, in the out-side-longitudinal system it can be derived that 
the correlation radii  are given by the space-time 
variances as
\begin{subequations}
\begin{eqnarray}
R_o^2(K) & = & \left \langle (\tilde x_o - \beta_o \tilde t)^2   \right \rangle   \\
R_s^2(K) & = & \left \langle x_s^2 \right \rangle \\
R_l^2(K) & = & \left \langle (\tilde x_l - \beta_l \tilde t)^2 \right \rangle \\
R_{os}^2(K) & = & \left \langle (\tilde x_o - \beta_o \tilde t)\tilde x_s\right \rangle\\
R_{ol}^2(K) & = & \left \langle (\tilde x_o - \beta_o \tilde t)(\tilde x_l - \beta_l \tilde t)\right \rangle \\
R_{sl}^2(K) & = & \left \langle \tilde x_s (\tilde x_l - \beta_l \tilde t) \right \rangle\,  .
\end{eqnarray}
\end{subequations}
Note that the space-time variances depend on average momentum $K$. 
Here we have introduced the averaging over the source 
\begin{equation}
\langle f(x) \rangle = 
\frac{\int S(x,p)\, f(x)\, d^4x}{\int S(x,p)\, d^4x}
\end{equation}
and we also introduced the shifted coordinates as
\begin{equation}
\tilde x^\mu = x^\mu - \langle x^\mu \rangle \, .
\end{equation}

Recall that the coordinates $x_o$, $x_s$, $x_l$ are connected with the direction of the emitted particles.
The explicit angular dependence is obtained simply by expressing the out-side-longitudinal 
coordinates in terms of the coordinates $x$, $y$, which are  fixed with the fireball:
\begin{subequations}
\begin{eqnarray}
\tilde x_o & = & \tilde x \cos\phi + \tilde y \sin\phi\\
\tilde x_s & = & -\tilde x \sin\phi + \tilde y \cos\phi\,  .
\end{eqnarray}
\end{subequations}
where $\phi$ is the azimuthal angle of the emitted hadron pairs. This leads to 
\begin{subequations}
\label{e:mie}
\begin{eqnarray}
\nonumber
R_s^2 & = & \frac{1}{2}\left ( \langle \tilde x^2 \rangle + \langle \tilde y^2 \rangle  \right )
+ \frac{1}{2} \left ( \langle \tilde x^2 \rangle - \langle \tilde y^2 \rangle  \right )\cos2\phi
\label{e:mie-a}
\\  && 
- \langle \tilde x \tilde y \rangle \sin2\phi\\
R_o^2 & = &  \frac{1}{2}\left(\left\langle \tilde{x}^2\right\rangle +\left\langle \tilde{y}^2\right\rangle \right) 
- \frac{1}{2}\left(\left\langle \tilde{y}^2\right\rangle -\left\langle \tilde{x}^2\right\rangle \right)\cos 2\phi 
\nonumber \\ && \nonumber
+ \left\langle \tilde{x}\tilde{y}\right\rangle \sin 2\phi  
+ \beta_o^2\left\langle \tilde{t}^2\right\rangle -2\beta_o\left\langle \tilde{x}\tilde{t}\right\rangle \cos \phi
\\ &&
 -2\beta_o \left\langle \tilde{y}\tilde{t}\right\rangle \sin \phi 
\\
R_l^2 & = & \left\langle \left( \tilde{z}-\beta_l \tilde{t}\right) ^2\right\rangle 
\\
R_{os}^2 & = & \left\langle \tilde{x}\tilde{y}\right\rangle \cos 2\phi 
+\frac{1}{2}\left( \left\langle \tilde{y}^2\right\rangle -\left\langle \tilde{x}^2\right\rangle \right) \sin 2\phi 
\nonumber \\ && 
+ \beta_o\left\langle \tilde{x}\tilde{t}\right\rangle \sin \phi 
-\beta_o \left\langle \tilde{y}\tilde{t}\right\rangle \cos \phi
\\
R_{ol}^2 & = & \left\langle \left( \tilde{z}-\beta_l \tilde{t}\right) \tilde{x}\right\rangle \cos \phi 
+\left\langle \left( \tilde{z}-\beta_l \tilde{t}\right) \tilde{y}\right\rangle \sin \phi 
\nonumber \\ && 
- \beta_l \left\langle \left( \tilde{z}-\beta_l \tilde{t}\right) \tilde{t}\right\rangle 
\\
R_{sl}^2 & = & \left\langle \left( \tilde{z}-\beta_l \tilde{t}\right) \tilde{y}\right\rangle \cos \phi -\left\langle \left( \tilde{z}-\beta_l \tilde{t}\right) \tilde{x}\right\rangle \sin \phi \,  .
\label{e:mie-f}
\end{eqnarray}
\end{subequations}

In what follows we want to study the azimuthal dependence of the correlation radii. To this end, 
they are customarily expanded into Fourier series. The space-time co-variances also depend
non-trivially on $\phi$, although we have suppressed writing this out explicitly. They can be written 
out and inserted in the right-hand sides of Eqs.~(\ref{e:mie}). Then, by combining all sine and cosine terms 
on the r.h.s.\ of the obtained equations 
%and using addition theorems for trigonometric functions
one can analytically identify all terms of Fourier expansion of the correlation radii. This has been 
done earlier for the second order \cite{Heinz:2002au,Tomasik:2002rx}. The results of the 
third and higher orders are presented in Appendix~\ref{a:fourord}.

In practical calculation, however, one can proceed differently. The whole azimuthal dependence
can be calculated from Eqs.~(\ref{e:mie-a})-(\ref{e:mie-f}). Then one can extract any Fourier coefficient from 
the result. With increasing complexity of higher order terms this procedure appears computationally
more efficient. 

Moreover, these are not yet the correlation radii which correspond to the measured ones, even if 
the assumption of Gaussian correlation function is valid. 
In order to measure the correlation radii in real collisions, hadron pairs must be collected over 
a large number of events. The events must be rotated so that the event planes are all aligned. 
Otherwise the azimuthal dependence would be averaged out. For measuring the second-order 
oscillations one rotates the events so as to align the second-order event planes. For the third-order 
oscillation one aligns the third-order event planes. Since the event planes of different orders 
%are uncorrelated, 
may be assumed to be uncorrelated
these alignments effectively introduce averaging over the direction of the other 
event planes. This must be included in calculations. It has been investigated in 
\cite{Lokos:2016fze} that this averaging may introduce a few percent effect on the resulting 
correlation radii. Hence, when calculating the correlation radii from Eqs.~(\ref{e:mie}),
one additional integral over $\theta_2$ or $\theta_3$ must be calculated, 
depending on which order of oscillations we shall be interested in. 
We include this averaging over other event planes in our calculation. 

Therefore, the second order and the third order Fourier amplitudes that we are going to 
calculate do not belong to the same Fourier series. In the former case the emission function 
is averaged over all possible values of $\theta_3$, in the latter averaging runs over $\theta_2$.
This will be also reflected in notation: the correlation radii will be expanded into series 
\begin{equation}
R_i^2(\phi) = R_{i,j,0}^2 + \sum_{n=1}^\infty R_{i,j,n}^2 \cos(n(\phi-\phi_n))
\end{equation}
where $i = o,s$, and $j=2,3$, depending on which event plane $\theta_j$ has been put to 0. 
In general, terms of the same order may differ, if they come from averaging with different event-planes 
fixed. For example, we note that 
\begin{subequations}
\begin{eqnarray}
R_{o,2,0}^2 & \ne & R_{o,3,0}^2  \\
R_{s,2,0}^2 & \ne & R_{s,3,0}^2\,  .
\end{eqnarray}
\end{subequations}

\begin{figure}
\begin{center}
\includegraphics[width=0.48\textwidth]{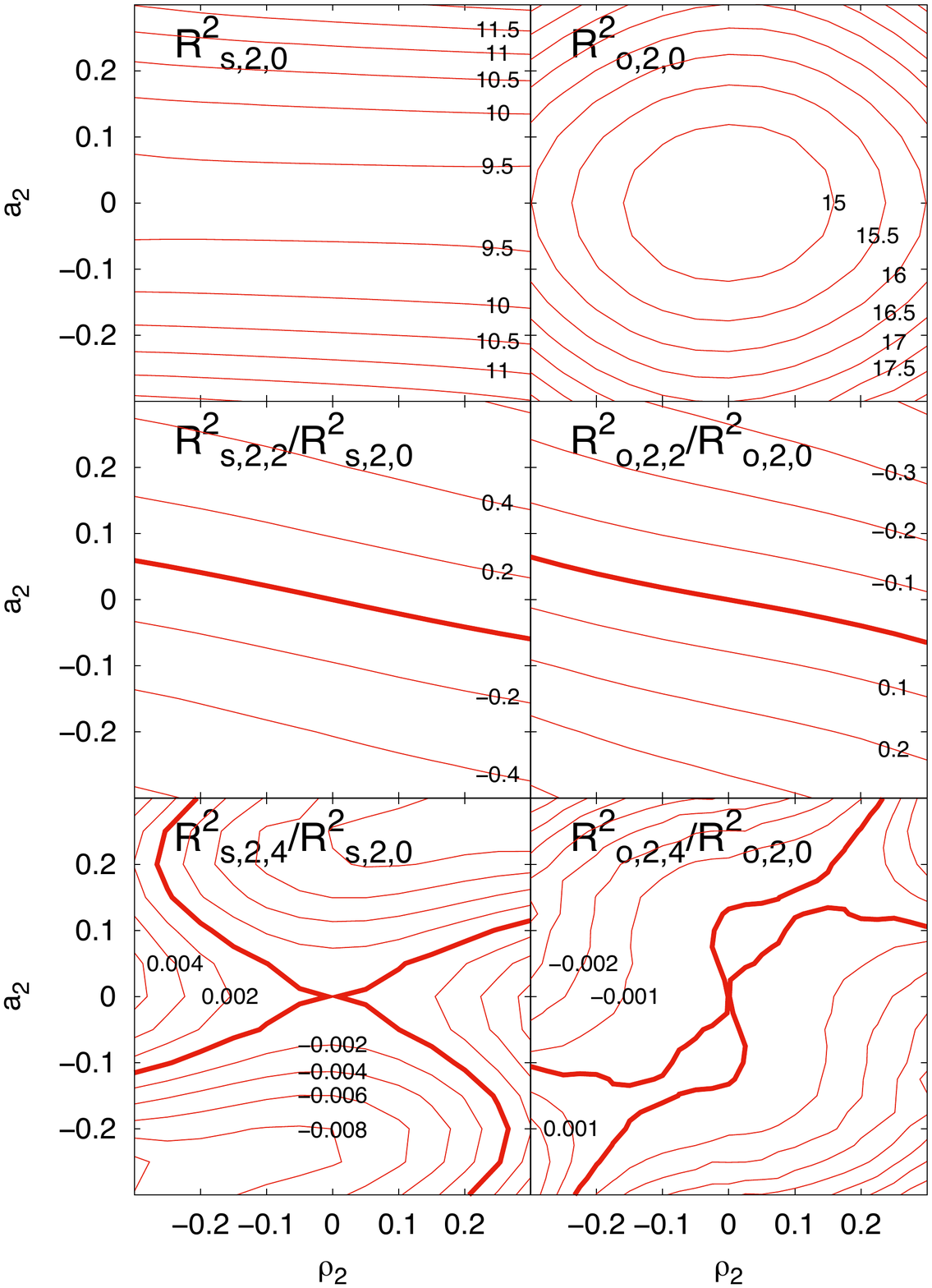}
\end{center}
\caption{
The dependence of transverse correlation radii and their oscillation amplitudes on 
second-order anisotropies in space and transverse flow. The values are calculated
for outward pair momentum $K_o  = 300$~MeV. Model parameters used in the calculation 
are $T=120$~MeV, $\rho_0 = 0.8$,
$R_0 = 7$~fm, $\tau_0 = 10$ fm/$c$. Third-order anisotropy parameters were set to 
$a_3 = \rho_3 = 0.1$. 
}
\label{f:R2osc}
\end{figure}

Second order oscillations have been calculated in \cite{Tomasik:2004bn}, but the averaging over 
third order event plane was not performed there. In order to fill this gap, we have done the 
calculation here and show the results in Fig.~\ref{f:R2osc}
%
%
%%%%%%%%%%%%%%%%%%%%%%
%%%%%%%%%%%%%%%%%%%%%%
%
The most important parameter which sets the scale of the transverse correlation radii is 
$R_0$.
We can see that even the average radii $R_{i,2,0}^2$ depend on both anisotropy parameters
$a_2$ and $\rho_2$. 
For the higher-order Fourier terms, we would like to factorise out their trivial scaling  
with $R_0$, thus for the analysis we divide all amplitudes by the zeroth-order term.
As it was observed previously \cite{Tomasik:2004bn}, the second order oscillation 
amplitude is mainly set by the spatial anisotropy parameter $a_2$. The dependence 
on $\rho_2$ is weak. This confirms the early conjecture that the second-order spatial 
deformation can be measured with the help of correlation radii \cite{Tomasik:2004bn}. 
Both spatial and flow anisotropy can then be obtained from combined measurement of 
$v_2$ and the azimuthal dependence of correlation radii. 
Figure~\ref{f:R2osc} also shows that in this model higher-order terms in the decomposition 
of the correlation radii are very small. The fourth-order terms are smaller than the second-order
terms by two orders of magnitude. Although it shows an interesting dependence on 
$a_2$ and $\rho_2$, it is most likely below any reasonable experimental sensitivity.

\begin{figure}
\begin{center}
\includegraphics[width=0.48\textwidth]{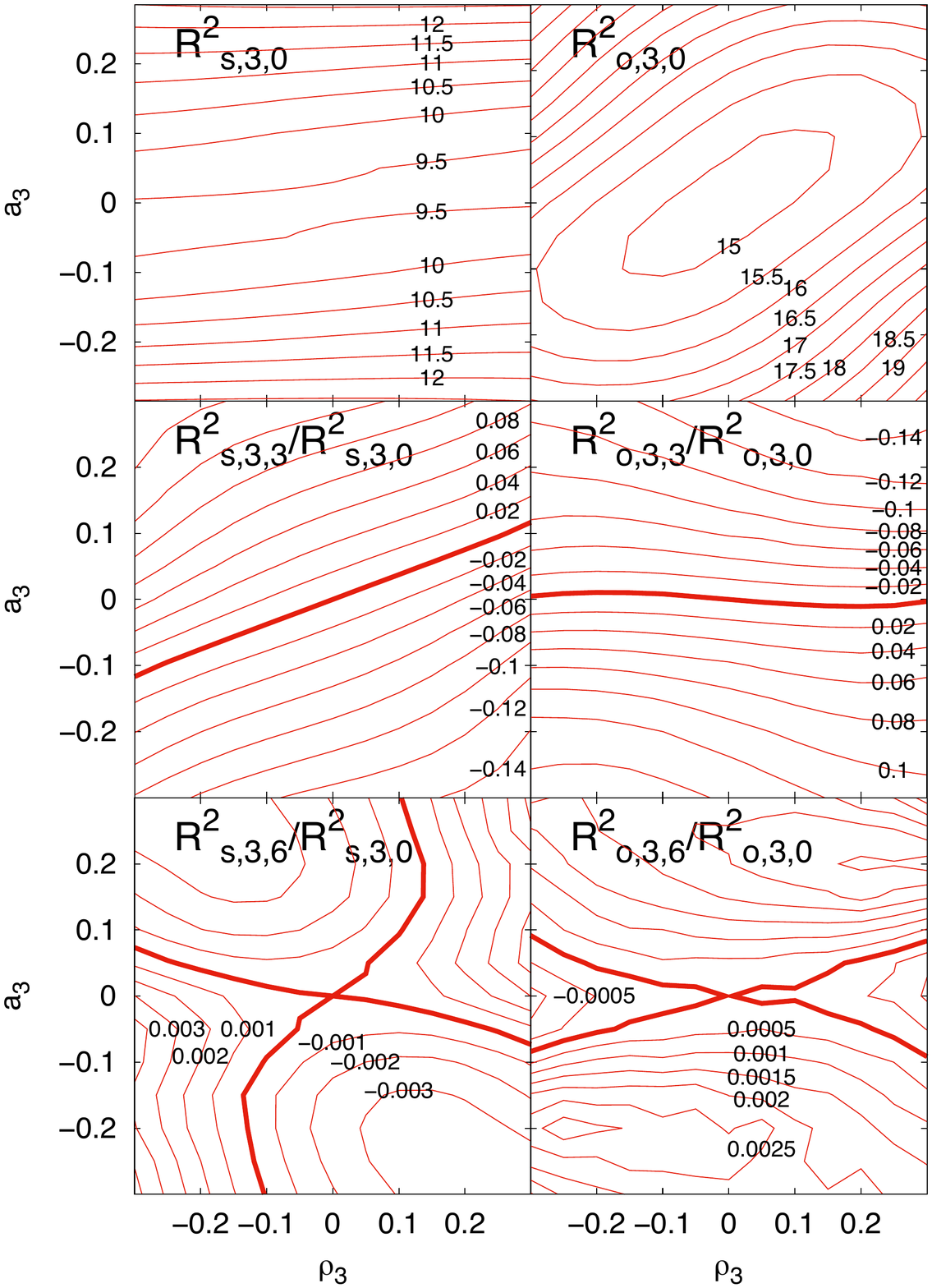}
\end{center}
\caption{
The dependence of transverse correlation radii and their oscillation amplitudes on 
third-order anisotropies in space and transverse flow. The values are calculated
for outward pair momentum $K_o  = 300$~MeV. Model parameters used in the calculation 
are $T=120$~MeV, $\rho_0 = 0.8$,
$R_0 = 7$~fm, $\tau_0 = 10$ fm/$c$. Second-order anisotropy parameters were set to 
$a_2 = \rho_2 = 0.1$. 
}
\label{f:R3osc}
\end{figure}

We have also looked at third-order oscillation in case of averaging over all possible directions 
of the second-order event plane. The resulting dependence of the correlation radii on $a_3$ 
and $\rho_3$ is plotted in Fig.~\ref{f:R3osc}.

%
%%%%%%%%%%%%%%%%%%%%%%
%%%%%%%%%%%%%%%%%%%%%%
%
Again, even the azimuthally averaged radii show some dependence on both anisotropy parameters. 
The third-order spatial anisotropy is best reflected in the third-order scaled amplitude of the 
outward radii $R_{o,3,3}^2/R_{o,3,0}^2$. For the sideward radius the third-order oscillation 
depends on both $a_3$ and $\rho_3$.  Note, however, that the third-order oscillation is typically 
smaller by an order of magnitude if it is compared to the second-order oscillation in 
Fig.~\ref{f:R2osc}. Even more suppressed is the next 
higher order, which is the sixth in this case. In absolute numbers 
the scaled amplitudes are on the level of a few per mille or even less. We do not
expect that such a weak signal could be reasonably measured in experiments.

%%%%%%%%%%%%%%%%%%%%%%%%%%%%%%%%%%%%%%%%%%%%%%%%%%%%%%

\section{Relation to data}
\label{s:data}

The present model has been designed with the aim to better characterise measured data on spectra and 
anisotropies. However, such an analysis requires to take into account many more issues and is technically 
much more involved. In order to reach physically relevant results, resonances must be included in the 
analysis \cite{Melo:2015wpa}. This highly increases the complexity of calculations. Then, all data, 
i.e.\ identified spectra, anisotropy coefficients, and correlations, should be fitted simultaneously. 
The new Bayesian technique \cite{Bernhard:2016tnd} seems well suitable to this aim. Such a thorough analysis,
however, goes far beyond our scope here. Nevertheless, we want to illustrate the qualitative features presented in 
previous sections with the help of comparison to data. We hasten to stress that this comparison should be understood
merely on  qualitative level. 

The STAR Collaboration has measured data from  Au+Au collisions at $\sqrt{s_{NN}} = 200$~GeV. They 
have analysed  second-order oscillations of correlation radii as functions of  azimuthal angle
with the blast-wave model extended to  that order in \cite{Adams:2004yc}. In that analysis, the temperature $T$ and 
transverse flow gradient $\rho_0$ were inferred from a simultaneous fit to pion, kaon, and proton $p_t$ spectra and $v_2$.
Analysis of the azimuthally sensitive correlation radii yielded the sizes of the fireball and the second-order anisotropy 
parameters. 
% Mate suggestion (slightly reformulated)
From this analysis of 10-20\% centrality STAR data, the following model parameter values have been extracted
by the STAR Collaboration \cite{Adams:2004yc}:
%Here,  we shall use STAR measurement in 10-20\% centrality class.
%This leads to the values which will be used here: 
$T = 98$~MeV, $\rho_0 =0.98$, $\rho_2 = 0.05$, $\tau_0 = 7.8$~fm/$c$,
$\Delta\tau = 2.59$~fm/$c$. The second-order spatial anisotropy can be translated into our model as $R_0 = 11.4$~fm,
$a_2 = 0.0439$.

\begin{figure}
\begin{center}
\includegraphics[width=0.48\textwidth]{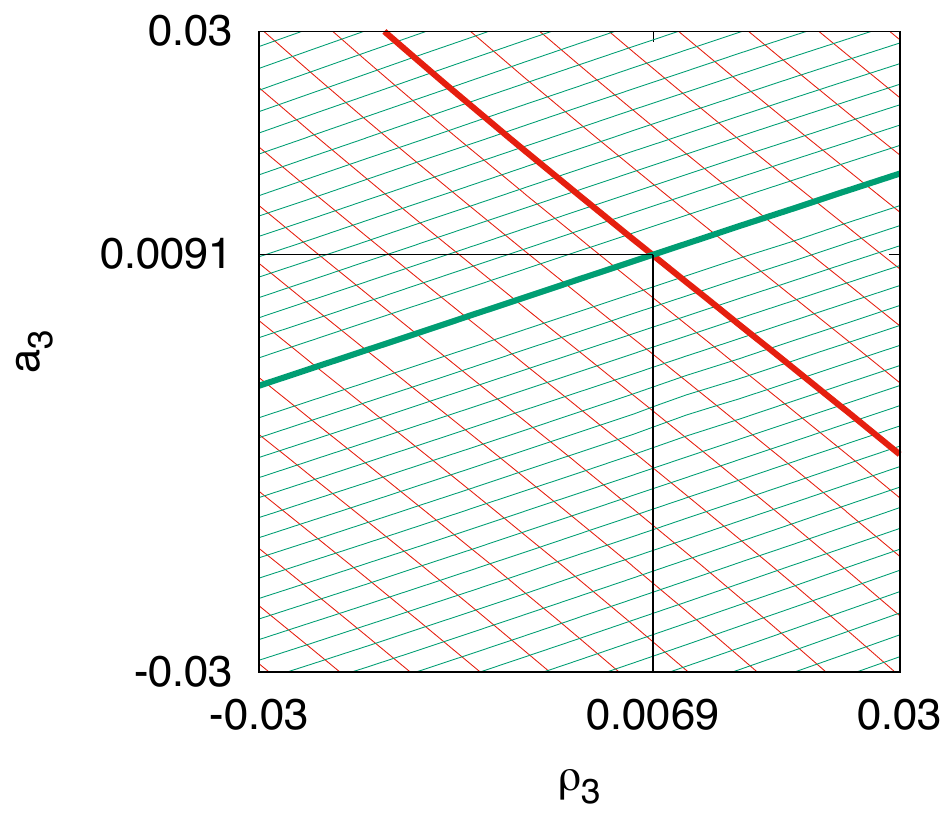}
\end{center}
\caption{Combined contour lines of constant $v_3$ (red from top left to bottom right) and constant 
$R_{s,3,3}^2/R_{s,0,3}^2$ (green from bottom left to top right). Thick lines show the values of data for 
$p_t = 863$~MeV ($v_3$) \cite{Adare:2011tg} and $K_t = 877$~MeV (correlation radii) \cite{Adare:2014vax}.
The increment between neighbouring lines is 0.01.
}
\label{f:3data}
\end{figure}

With these values fixed we calculate the combined contour plot for the dependences of $v_3$ and third-order scaled amplitude 
$R_{s,3,3}^2/R_{s,3,0}^2$ on $a_3$ and $\rho_3$. It is shown in Fig.~\ref{f:3data}.
%%%%%%%%%%%%%%%%%%%%%%
%%%%%%%%%%%%%%%%%%%%%%
The curves decreasing to the right are lines with constant $v_3$, the others correspond to constant 
$R_{s,3,3}^2/R_{s,0,3}^2$. Thick lines
represent the data values by PHENIX \cite{Adare:2011tg,Adare:2014vax}. 
Out of available data we had to choose bins in $p_t$ for $v_3$ and correlation radii, 
which overlap. Unfortunately, these two sets of data are  measured in slightly different momentum ranges. 
In order to use overlapping $p_t$ bins, we have taken $v_3$ point for pions at $p_t = 863$~MeV
\cite{Adare:2011tg} and correlation radii for $K_t = 877$~MeV \cite{Adare:2014vax}. 
The values extracted from such a simple 
comparison with data are: $a_3 = 0.0091$ and $\rho_3 = 0.0069$. 

\begin{figure}
\begin{center}
\includegraphics[width=0.48\textwidth]{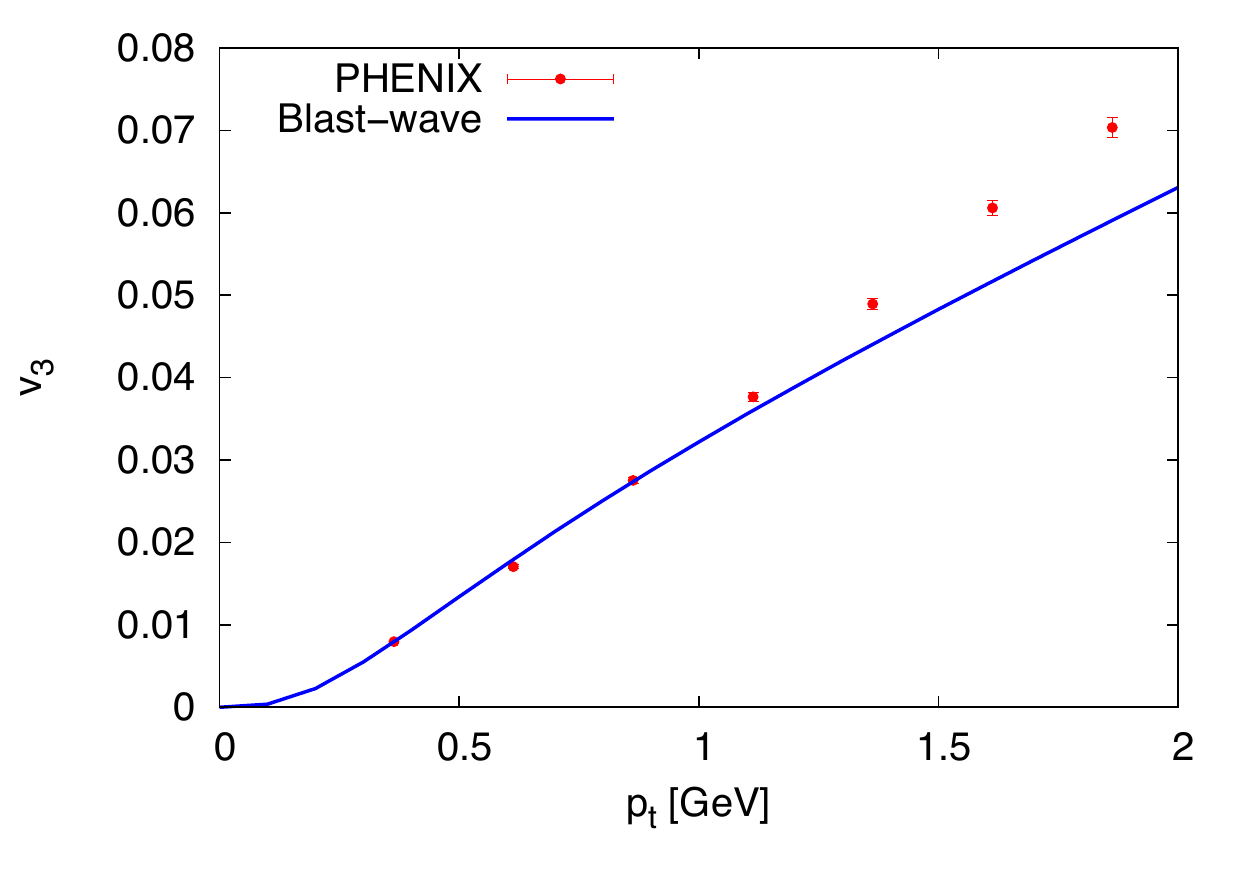}
\end{center}
\caption{The $p_t$ dependence of $v_3$ for identified pions measured by PHENIX collaboration 
\cite{Adare:2011tg} compared with the theoretical curve from the blast-wave model with parameters
determined in Fig.~\ref{f:3data}.
}
\label{f:v3data}
\end{figure}

\begin{figure}
\begin{center}
\includegraphics[width=0.42\textwidth]{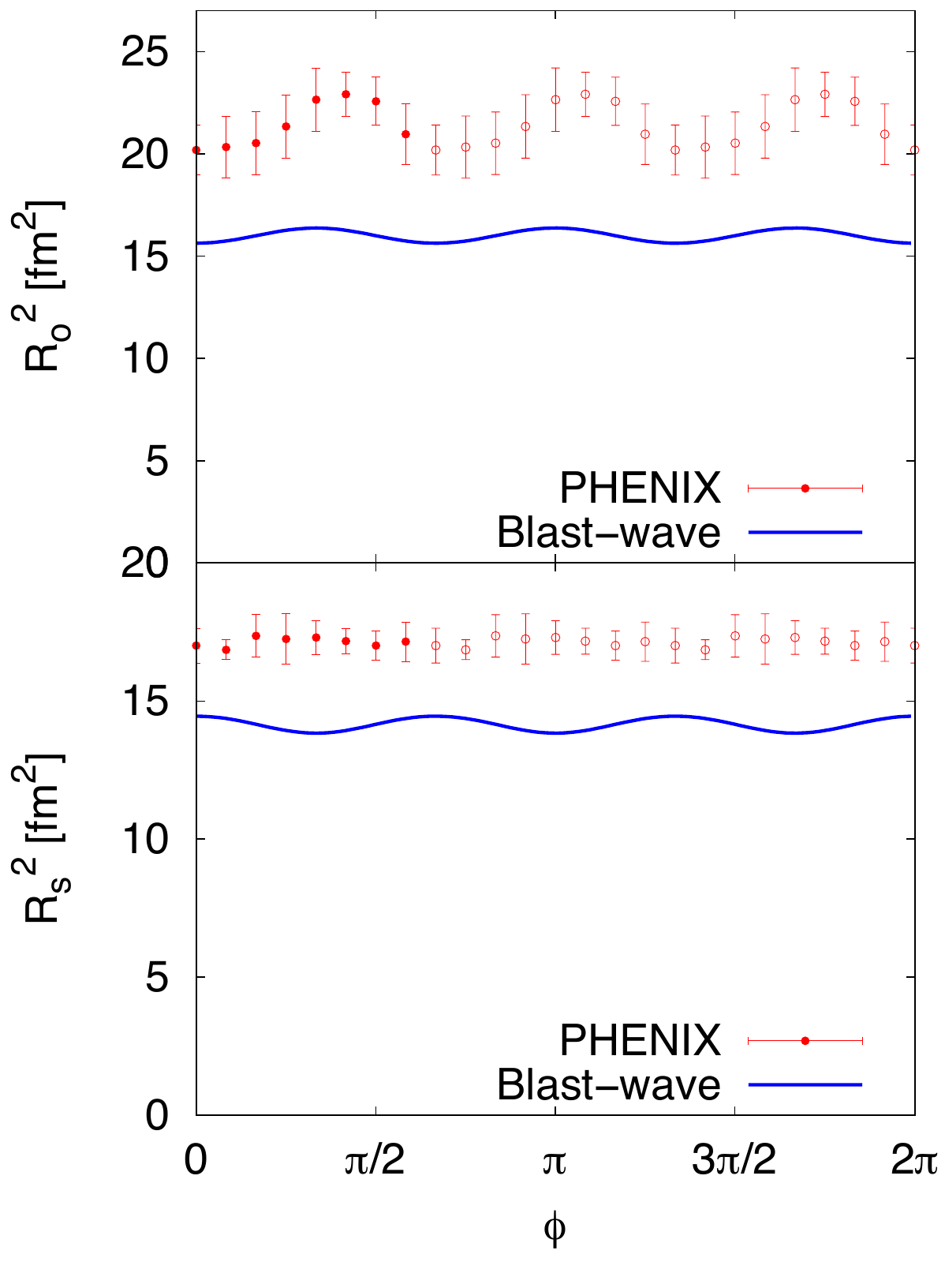}
\end{center}
\caption{The azimuthal dependence of correlation radii with respect to the third-order event 
plane for $K_t = 530$~MeV/$c$.
Data by PHENIX collaboration \cite{Adare:2014vax} are compared with results from the blast-wave model 
with parameters determined in Fig.~\ref{f:3data}.
}
\label{f:asHBTdata}
\end{figure}

With the extracted model parameters we tried to calculate theoretical predictions for $v_3$ 
(Fig.~\ref{f:v3data})
and the azimuthal angle dependence of the correlation radii (Fig.~\ref{f:asHBTdata}).
%%%%%%%%%%%%%%%%%%%%%%
%%%%%%%%%%%%%%%%%%%%%%
%%%%%%%%%%%%%%%%%%%%%%
%%%%%%%%%%%%%%%%%%%%%%
For the latter we chose to plot the radii measured for $K_t = 530$ MeV/$c$ as was also done by the PHENIX 
Collaboration in \cite{Adare:2014vax}.
The model fails completely in reproducing the absolute size of the correlation radii. The corresponding parameter, 
however, has been fixed from the STAR analysis of second-order oscillations of correlation radii. The mean 
values of the radii should be very similar for second and the third order oscillations. This clearly 
demonstrates the need of simultaneous fit to all available data in single analysis if one wants to go beyond the 
qualitative level. 

\begin{figure}
\begin{center}
\includegraphics[width=0.42\textwidth]{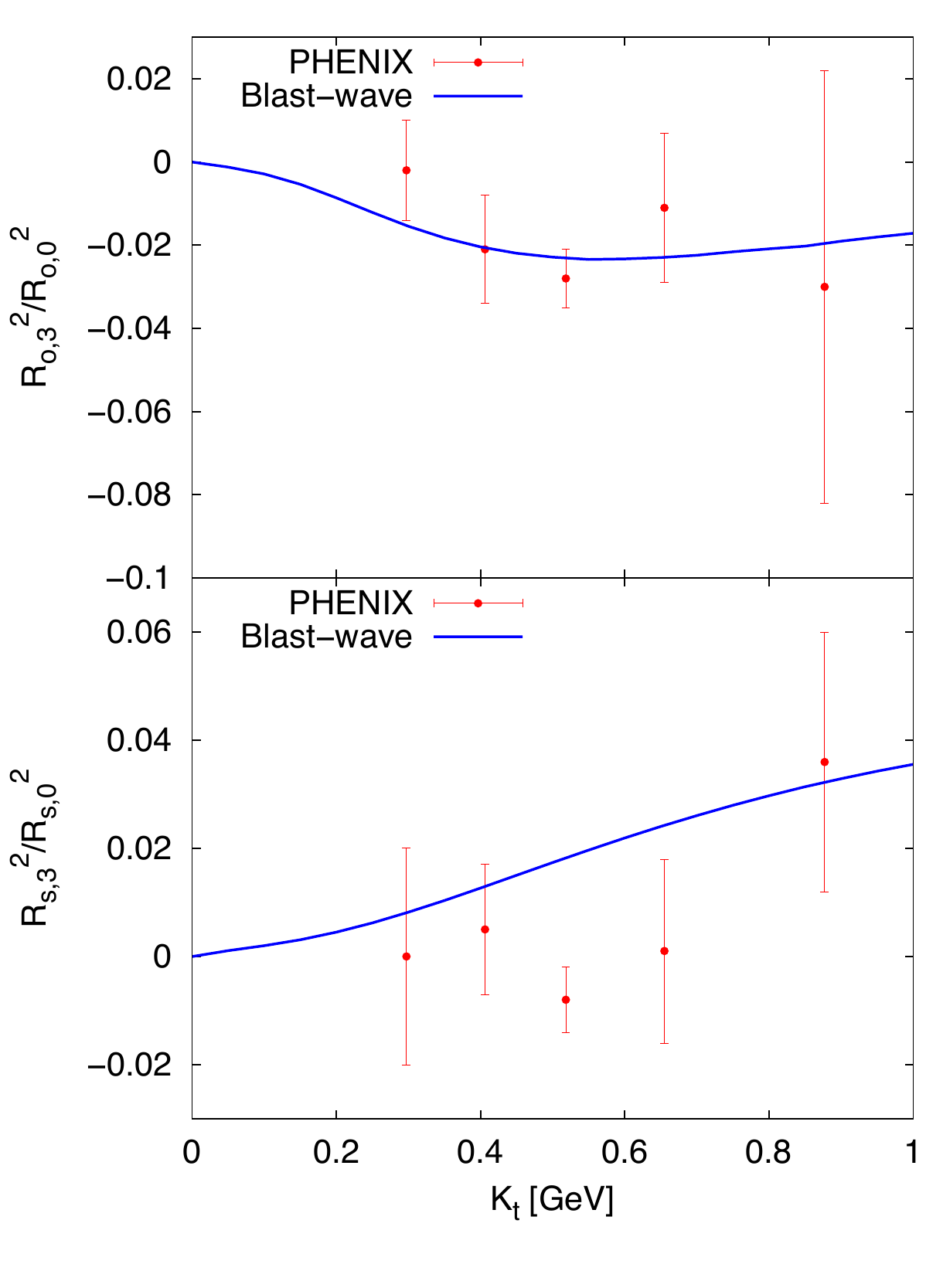}
\end{center}
\caption{The $K_t$ dependence of third-order oscillation amplitude of the correlation radii.
Data by PHENIX collaboration \cite{Adare:2014vax} are compared with results from the blast-wave model 
with parameters determined in Fig.~\ref{f:3data}.
}
\label{f:c3aml}
\end{figure}

We also show in Fig.~\ref{f:c3aml} the third-order scaled amplitudes of correlation radii as functions of $K_t$. 
%%%%%%%%%%%%%%%%%%%%%%
%%%%%%%%%%%%%%%%%%%%%%
The experimental error bars are huge and we explained that we have based our analysis on the point at highest 
$K_t$. It seems that the PHENIX data require amplitudes of opposite signs, which the blast-wave model can 
accommodate. 

We want to close this section with a few comments on the applicability of the model to 
preliminary ALICE data from Pb+Pb collisions at $\sqrt{s_{NN}} = 2.76$~TeV \cite{ALICE@WPCF}.
The data seem to indicate that the third-order oscillation amplitudes of both $R_s^2$ and $R_o^2$
are negative. By inspecting Fig.~\ref{f:R3osc} we find that such situation only happens in a 
small region in parameter space with $a_3$ around 0 and positive $\rho_3$. Thus from combined 
measurements of outward and sideward radii one could deduce that the fireball at the LHC has rather symmetric 
shape and the anisotropy is set by the transverse collective velocity field.

%%%%%%%%%%%%%%%%%%%%%%%%%%%%%%%%%%%%%%%%%%%%%%%%%%%%%%

\section{Conclusions}
\label{s:conc}

We have generalised the blast-wave model so that it includes third-order anisotropies in both space and 
expansion. Analogically to the second order, from the combination of Fig.~\ref{f:v2cntr}
and Fig.~\ref{f:R3osc} we can infer, that it is indeed possible to reconstruct both anisotropy 
coefficients of this model: $a_3$ and $\rho_3$, from measurements of the azimuthal anisotropy 
of single-particle momentum distributions and HBT radii.

We have also pointed out the need for averaging over the difference of second and third-order 
reaction planes when focusing on a selected order of Fourier decomposition  of the hadron distribution 
or the correlation radii. This is effectively done in data analysis when all events are aligned according to 
the event plane of the selected order. 

The contour plots shown in Fig.~\ref{f:R3osc} exemplify the statement made in \cite{Plumberg:2013nga}
that at fixed flow anisotropy the 
amplitude of correlation radii oscillation can be tuned with the help of spatial anisotropy and even a flip 
in the phase can be obtained. Such a flip of the phase corresponds to a change of the sign of the 
amplitude. Keeping constant flow anisotropy and changing space anisotropy corresponds to moving 
vertically in the panels of Fig.~\ref{f:R3osc} and the phase flip corresponds to crossing the thick line 
in that Figure. Our results are much more detailed since we show the full dependence on the two parameters. 

We have also calculated the subleading terms for the third-order anisotropies in the oscillations of correlation radii. 
These are the sixth-order oscillations. They were shown to be an order of magnitude  smaller than the lower order
and hardly measurable at current experimental statistics. We have actually derived expressions for oscillation amplitudes
at general order, but there is currently no need to go to higher orders also with model studies as sufficient
statistics would hardly be available.

%
%%%%%%%%%%%%%%%%%%%%%%
\begin{figure}
\begin{center}
\includegraphics[width=0.48\textwidth]{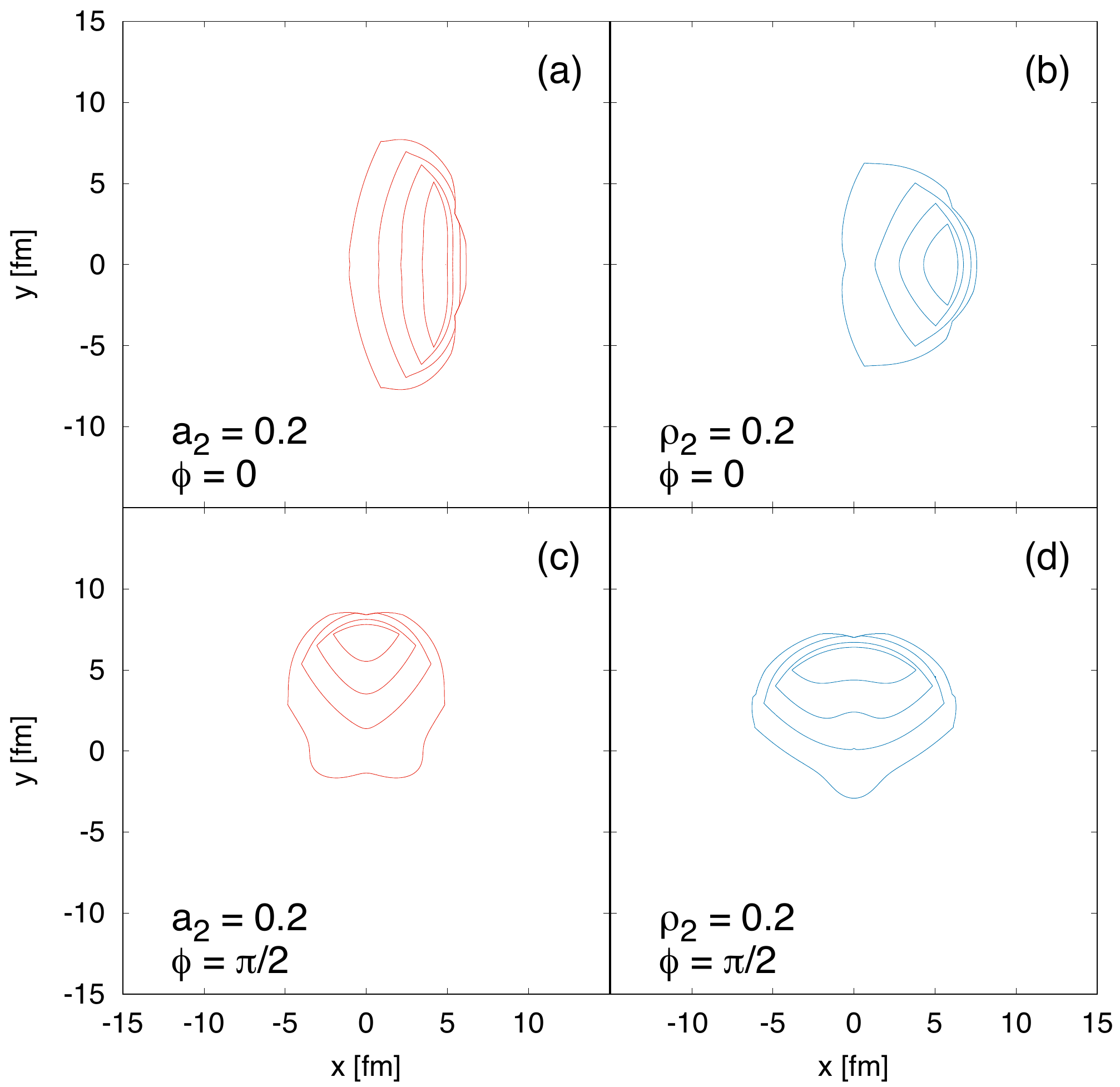}
\end{center}
\caption{
Transverse profiles of the emission function for the blast-wave model  with 
source parameters as in Fig.~\ref{f:R3osc}, for particles with $p_t = 300$~MeV
in the indicated direction $\phi$. Left column: only density profile anisotropy; right column: only flow anisotropy. 
Other source parameters are  $T=120$~MeV, $\tau_0=10$~fm, $R_0=7$~fm, $\Delta \tau=1$~fm, $\rho_0=0.8$.
Shown are the effective sources which are calculated by integrating over all directions of 
the third-order anisotropy with $a_3 = \rho_3 = 0.1$. 
The contours correspond to levels of 0.8, 0.6, 0.4, and 0.2  of the maximum. 
}
\label{f:spBW}
\end{figure}
%%%%%%%%%%%%%%%%%%%%%%
%
%
%%%%%%%%%%%%%%%%%%%%%%
\begin{figure}
\begin{center}
\includegraphics[width=0.48\textwidth]{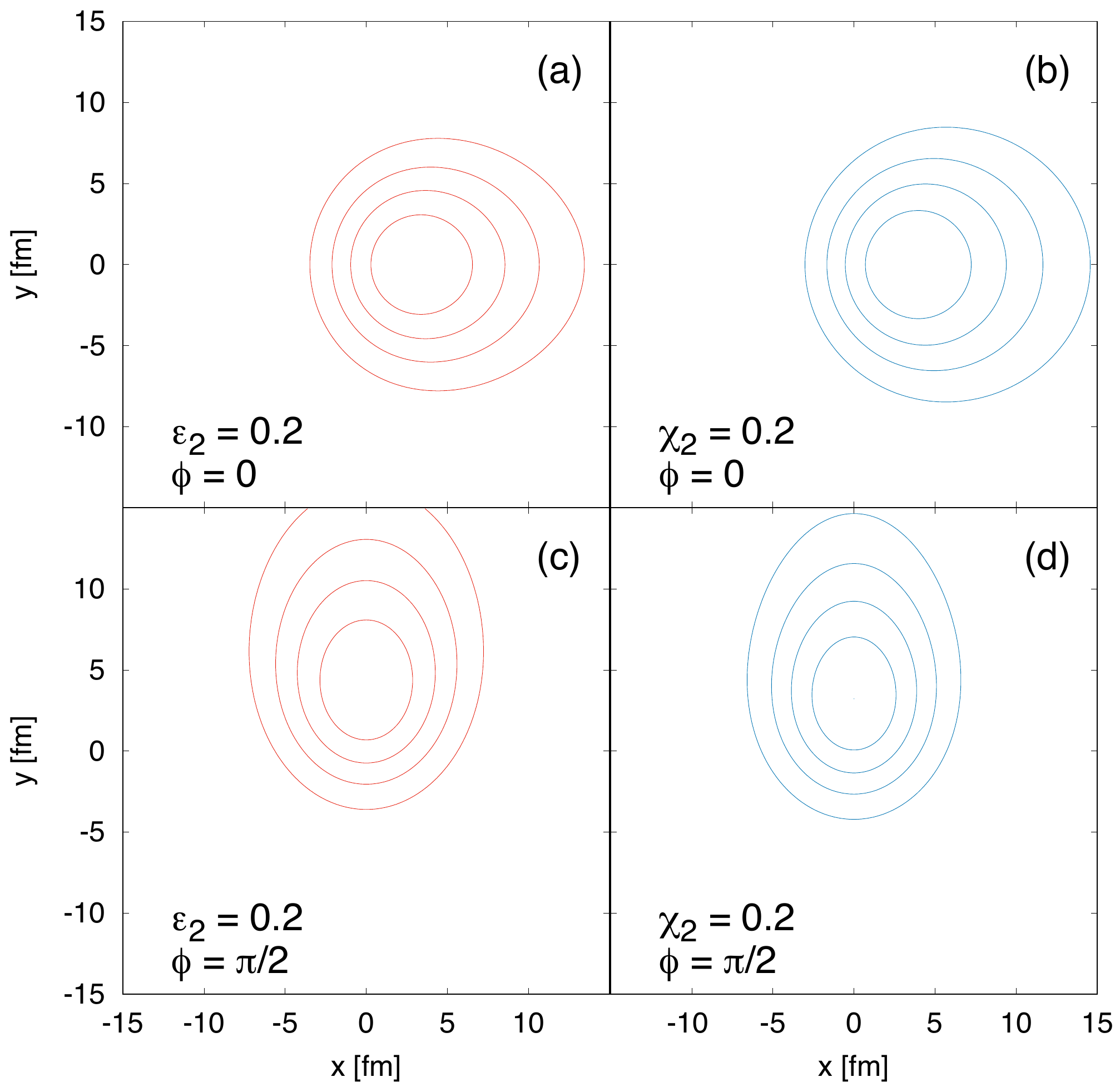}
\end{center}
\caption{
Analogical to Fig.~\ref{f:spBW}, but calculated with the Buda-Lund model, with source parameters
$T=120$~MeV, $\tau_0 = 10$~fm, $a_2 = 0.3$, $b=-0.1$, $R=7$, $Z=15$, $H=8$, $H_z=16$,  
$\epsilon_3 = \chi_3 = 0.1$. The model is explained in \cite{Lokos:2016fze}. The values 
of the second-order anisotropy parameters are indicated in the figure. The averaging over all 
directions of the third-order anisotropy is performed with $\varepsilon_3 = \chi_3 =0.1$.
}
\label{f:spBL}
\end{figure}
%%%%%%%%%%%%%%%%%%%%%%
%
It is interesting to compare our results to those obtained in an analogical study which used the 
Buda-Lund model \cite{Lokos:2016fze}. In that model, the way in which observables depend on 
the combination of the space and flow anisotropy is different to the one presented here. 
Partially, this is due to our definition of the spatial profile with the minus sign in 
Eq.~(\ref{e:Rtheta}). The corresponding sign in the Buda-Lund model was kept to be plus. 
On the qualitative level, the oscillations of the correlation radii within the Buda-Lund model 
seem to be much more sensitive to the flow anisotropy than it is the case in this study. 

In order to explore the difference of the two models in more details, in Figs.~\ref{f:spBW} and 
\ref{f:spBL} we plot transverse profiles of the emitting sources according to the blast-wave 
and Buda-Lund models, respectively. Plotted are not the emission functions directly. We have 
assumed that there are the third-order anisotropies, as well, which have their third-order event
planes completely independent from the second-order event plane directions. Then, we have 
integrated over all possible directions of the third-order event plane and obtained an 
\emph{effective} emission function with only second-order anisotropy. 
In the figures, we have assumed pions with $p_t = 300$ MeV and two different azimuthal
angles of particle emission: $\phi = 0$ (upper rows) and $\phi = \pi/2$ (lower rows). 

Figures \ref{f:spBW} and \ref{f:spBL} clearly demonstrate that the two models behave 
rather differently. In the Buda-Lund model, the deformations caused by the density anisotropy
(Fig.~\ref{f:spBL} left) and by flow anisotropy (Fig.~\ref{f:spBL} right) look 
qualitatively similar. On the other hand, the sharp source boundary in the blast wave 
model has severe influence. The spatial anisotropy (Fig.~\ref{f:spBW} left) makes the two 
sources for $\phi = 0$ and $\phi = \pi/2$ qualitatively much more different from 
each other than the flow anisotropy (Fig.~\ref{f:spBW} right). Note that the sources should 
be compared in the out-side-long coordinate system, and the outward axis is directed in the 
direction of $\phi$. Thus we see the much stronger dependence on shape anisotropy 
than on flow anisotropy. 

Having the two models which appear so differently, 
the choice of the more suitable model should be decided by data. Note, however, that neither 
this schematic study, nor the one of ref.~\cite{Lokos:2016fze} included meson production from 
the decays of resonances and the modification of the correlation function due to viscosity.   
Their influence should be investigated in the future 
in order to arrive at a conclusive answer.

%%%%%%%%%%%%%%%%%%%%%%%%%%%%%%%%%%%%%%%%%%%%%%%%%

\begin{acknowledgement}
This work has been performed in framework of COST Action CA15213 ``Theory of hot matter and 
relativistic heavy-ion collisions'' (THOR). 
Partial support by VEGA 1/0469/15 (Slovakia) and by M\v{S}MT grant No.~LG15001 (Czech Republic) is acknowledged.
M. Cs. was supported by the J\'anos Bolyai Research Scholarship of the Hungarian Academy of Sciences.
\end{acknowledgement}

%%%%%%%%%%%%%%%%%%%%%%%%%%%%%%%%%%%%%%%%%%%%%%%%%
\appendix
%%%%%%%%%%%%%%%%%%%%%%%%%%%%%%%%%%%%%%%%%%%%%%%%%

\section{Relation between different parametrisations of second-order spatial anisotropy}
\label{a:a1}

In this Appendix we derive the relation between our pa\-ra\-me\-tri\-sa\-tion of the transverse shape 
and the one used in \cite{Tomasik:2004bn}.

First of all, it should be clearly stated that the two parametrisations are different. Hence, the elliptic
shape that  has been used in \cite{Tomasik:2004bn} would be fully reproduced with the help of 
parametrisation (\ref{e:Rtheta}) only if higher-order terms are included. Of course, the importance
of higher orders drops with $n$. 

We shall assume here that the second-order anisotropy is small, i.e., the parameter $a$ from 
Eq.~(\ref{e:aparam}) is close to 1. The ellipse of \cite{Tomasik:2004bn} includes points with 
coordinates
\begin{subequations}
\begin{eqnarray}
r_x & = & R_x \cos \theta = a R' \cos\theta\\
r_y & = & R_y \sin\theta = \frac{R'}{a} \sin\theta\,  .
\end{eqnarray}
\end{subequations}
Thus the radius as function of the azimuthal angle is 
\begin{equation}
R = \sqrt{r_x^2 + r_y^2} = \sqrt{
a^2 {R'}^2 \cos^2\theta + \frac{{R'}^2}{a^2} \sin^2 \theta
}\,  .
\end{equation}
This can be rewritten as 
\begin{equation}
R = R' \sqrt{\frac{a^4 + 1}{2a^2}} 
\left ( 
1 + \frac{a^4 - 1}{a^4 + 1} \cos(2\theta) 
\right )^{\frac{1}{2}}\,  .
\end{equation}
Now we assume that $a\to 1$ and therefore $(a^4-1)/(a^4+1)$ is very small. Thus we can 
Taylor-expand the bracket up to first order and obtain
\begin{equation}
R(\theta) \approx R' \sqrt{\frac{a^4 + 1}{2a^2}} 
\left (
1 + \frac{1}{2} \frac{a^4 - 1}{a^4 + 1} \cos(2\theta) 
\right ) \,  .
\end{equation}

The mean radius $R_0$ of the present model is to be identified with 
\begin{equation}
R_0 = R' \sqrt{\frac{a^4 + 1}{2a^2}}\, .
\end{equation}

The amplitude of the oscillations is to be identified with $(-a_2)$
\begin{equation}
a_2 = - \frac{1}{2} \frac{a^4 - 1}{a^4 + 1}\,  .
\end{equation}
Inverting this relation gives
\begin{equation}
a = \left (
\frac{1-2a_2}{1+2a_2}
\right )^{\frac{1}{4}}\,  .
\end{equation}

%%%%%%%%%%%%%%%%%%%%%%%%%%%%%%%%%%%%%%%%%%%%%%%%%%

\section{The direction of transverse velocity}
\label{a:a2}

Here we derive the direction of the transverse velocity, which is given by the angle $\theta_b$. 
Since it is supposed to be perpendicular to the surface of constant $\rb$, we write down 
its coordinates
\begin{subequations}
\label{e:xyrb}
\begin{eqnarray}
x & = & \rb R_0 \left ( 1 - \sum_{n=2}^\infty a_n \cos(n(\theta - \theta_n))\right ) \cos\theta\\
y & = & \rb R_0  \left ( 1 - \sum_{n=2}^\infty a_n \cos(n(\theta - \theta_n))\right ) \sin\theta\, .
\end{eqnarray}
\end{subequations}
In further calculations, however, we shall truncate the expansion after the third-order term. 

The (truncated) expressions (\ref{e:xyrb}) can be inserted into the derivatives in Eq.~(\ref{e:thetab}).
This gives
\begin{equation}
\theta_b = \frac{\pi}{2} + \arctan
\frac{A}{B}
\,  .
\end{equation}
where 
\begin{multline}
A = 
4a_2\sin\theta + 3 a_3 \frac{\sin3(\theta-\theta_3)}{\cos\theta} \\- 
\frac{-1 + a_2\cos2(\theta-\theta_2) + a_3 \cos3(\theta-\theta_3)}{\sin\theta}
\end{multline}
\begin{multline}
B = 
a_2\cos\theta + 3 a_3 \frac{\sin3(\theta-\theta_3)}{\sin\theta} \\- 
\frac{-1 + a_2\cos2(\theta-\theta_2) + a_3 \cos3(\theta-\theta_3)}{\cos\theta}
\end{multline}

%%%%%%%%%%%%%%%%%%%%%%%%%%%%%%%%%%%%%%%%%%%%%%%%

\section{Fourier amplitudes of the correlation radii}
\label{a:fourord}

Here we give an overview of the Fourier amplitudes of the azimuthal dependence of 
outward, sideward, and out-side cross-term correlation radii. 
Note that these formulas are model-independent. 
Dependence on a particular model comes into evaluation of individual space-time 
(co-)variances. 
Note that analogical relations have been derived in \cite{Plumberg:2013nga}, where 
Milne coordinates were used instead of the Cartesian ones.

We define the amplitudes
\begin{subequations}
\begin{eqnarray}
R_o^2 & = & \left ( R_o^2 \right )_0 + \sum_{n=1}^\infty 
\left [
\left ( R_o^2 \right )_n^s \sin(n\phi) + \left ( R_o^2 \right )_n^c \cos(n\phi) 
\right ]
\nonumber  \\ && \\
R_s^2 & = & \left ( R_s^2 \right )_0 + \sum_{n=1}^\infty 
\left [
\left ( R_s^2 \right )_n^s \sin(n\phi) + \left ( R_s^2 \right )_n^c \cos(n\phi) 
\right ]
\nonumber  \\ && \\
R_{os}^2 & = & \left ( R_{os}^2 \right )_0 + \sum_{n=1}^\infty 
\left [
\left ( R_{os}^2 \right )_n^s \sin(n\phi) + \left ( R_{os}^2 \right )_n^c \cos(n\phi) 
\right ]
\nonumber  \\ && 
\end{eqnarray}
\end{subequations}
In a similar way we shall expand the space-time (co-)\-va\-rian\-ces
\begin{multline}
\label{e:covarexp}
\left \langle \tilde x^\mu \tilde x^\nu \right \rangle = 
\left \langle \tilde x^\mu \tilde x^\nu \right \rangle_0 
\\
+
\sum_{n=1}^{\infty} \left [
\left \langle \tilde x^\mu \tilde x^\nu \right \rangle_n^s \sin(n\phi)
+
\left \langle \tilde x^\mu \tilde x^\nu \right \rangle_n^c \cos(n\phi)
\right ]
\end{multline}

The series (\ref{e:covarexp}) are inserted into the model-independent expressions
(\ref{e:mie}) and the sine and cosine terms are reorganised with the help of addition 
theorems. Then one can collect the terms order by order and put them in equality with the 
corresponding expansion of the correlation radii. We obtain for the outward radius
\allowdisplaybreaks[1]
\begin{subequations}
\begin{eqnarray}
(R_o^2)_0 & = &  
\frac{1}{2}\left\langle \tilde{x}^2\right\rangle _0+\frac{1}{2}\left\langle \tilde{y}^2\right\rangle _0
-\frac{1}{4}\left\langle \tilde{y}^2\right\rangle _2^c
+\frac{1}{4}\left\langle \tilde{x}^2\right\rangle _2^c
\nonumber \\ &&  
+\frac{1}{2}\left\langle \tilde{x}\tilde{y}\right\rangle _2^s
+ \beta_o^2\left\langle \tilde{t}^2\right\rangle_0 
- \beta_o\left\langle \tilde{t}\tilde{x}\right\rangle_1^c 
\nonumber \\ &&
- \beta_o\left\langle \tilde{t}\tilde{y}\right\rangle_1^s 
\\
(R_o^2)_1^c & = & \frac{3}{4}\left\langle \tilde{x}^2\right\rangle _1^c
+\frac{1}{4}\left\langle \tilde{y}^2\right\rangle _1^c
-\frac{1}{4}\left\langle \tilde{y}^2\right\rangle _3^c
+\frac{1}{4}\left\langle \tilde{x}^2\right\rangle _3^c
\nonumber \\ &&
+\frac{1}{2}\left(\left\langle \tilde{x}\tilde{y}\right\rangle _1^s+\left\langle \tilde{x}\tilde{y}\right\rangle _3^s\right) 
+ \beta_o^2\left\langle \tilde{t}^2\right\rangle_1^c 
- 2\beta_o\left\langle \tilde{t}\tilde{x}\right\rangle_0 
\nonumber \\ &&
- \beta_o\left\langle \tilde{t}\tilde{x}\right\rangle_2^c - \beta_o\left\langle \tilde{t}\tilde{y}\right\rangle_2^s 
\\
(R_o^2)_1^s & = & 
\frac{1}{4}\left\langle \tilde{x}^2\right\rangle _1^s
+\frac{3}{4}\left\langle \tilde{y}^2\right\rangle _1^s
+\frac{1}{4}\left\langle \tilde{y}^2\right\rangle _3^s
-\frac{1}{4}\left\langle \tilde{x}^2\right\rangle _3^s
\nonumber \\ &&
+\frac{1}{2}\left(\left\langle \tilde{x}\tilde{y}\right\rangle _1^c-\left\langle \tilde{x}\tilde{y}\right\rangle _3^c\right) 
+\beta_o^2\left\langle \tilde{t}^2\right\rangle_1^s 
- \beta_o\left\langle \tilde{t}\tilde{x}\right\rangle_2^s 
\nonumber \\ &&
- 2\beta_o\left\langle \tilde{t}\tilde{y}\right\rangle_0 + \beta_o\left\langle \tilde{t}\tilde{y}\right\rangle_2^c
\\
(R_o^2)_2^c & = & 
\frac{1}{2}\left\langle \tilde{x}^2\right\rangle _2^c
+\frac{1}{2}\left\langle \tilde{y}^2\right\rangle _2^c
-\frac{1}{2}\left\langle \tilde{y}^2\right\rangle _0
-\frac{1}{4}\left\langle \tilde{y}^2\right\rangle _4^c
\nonumber \\ && 
+\frac{1}{2}\left\langle \tilde{x}^2\right\rangle _0
+\frac{1}{4}\left\langle \tilde{x}^2\right\rangle_4^c
+\frac{1}{2}\left\langle \tilde{x}\tilde{y}\right\rangle _4^s 
+ \beta_o^2\left\langle \tilde{t}^2\right\rangle_2^c 
\nonumber \\  && 
- \beta_o\left\langle \tilde{t}\tilde{x}\right\rangle_3^c 
- \beta_o\left\langle \tilde{t}\tilde{x}\right\rangle_1^c 
- \beta_o\left\langle \tilde{t}\tilde{y}\right\rangle_3^s 
\nonumber \\ &&
+\beta_o\left\langle \tilde{t}\tilde{y}\right\rangle_1^s
\\
(R_o^2)_2^s & = & 
\frac{1}{2}\left\langle \tilde{x}^2\right\rangle _2^s
+\frac{1}{2}\left\langle \tilde{y}^2\right\rangle _2^s
+\frac{1}{4}\left\langle \tilde{y}^2\right\rangle _4^s
-\frac{1}{4}\left\langle \tilde{x}^2\right\rangle_4^s
\nonumber \\ &&
+\left\langle \tilde{x}\tilde{y}\right\rangle _0
-\frac{1}{2}\left\langle \tilde{x}\tilde{y}\right\rangle _4^c 
+ \beta_o^2\left\langle \tilde{t}^2\right\rangle_2^s 
- \beta_o\left\langle \tilde{t}\tilde{x}\right\rangle_3^s 
\nonumber  \\ &&
- \beta_o\left\langle \tilde{t}\tilde{x}\right\rangle_1^s 
- \beta_o\left\langle \tilde{t}\tilde{y}\right\rangle_1^c 
+\beta_o\left\langle \tilde{t}\tilde{y}\right\rangle_3^c
\\
(R_o^2)_n^c & = & 
\frac{1}{2}\left\langle \tilde{x}^2\right\rangle _n^c
+\frac{1}{2}\left\langle \tilde{y}^2\right\rangle _n^c
-\frac{1}{4}\left\langle \tilde{y}^2\right\rangle _{n-2}^c
-\frac{1}{4}\left\langle \tilde{y}^2\right\rangle _{n+2}^c
\nonumber \\ && 
+\frac{1}{4}\left\langle \tilde{x}^2\right\rangle _{n-2}^c
+\frac{1}{4}\left\langle \tilde{x}^2\right\rangle _{n+2}^c 
+\frac{1}{2}\left\langle \tilde{x}\tilde{y}\right\rangle _{n+2}^s
\nonumber \\ &&
-\frac{1}{2}\left\langle \tilde{x}\tilde{y}\right\rangle _{n-2}^s 
+ \beta_o^2\left\langle \tilde{t}^2\right\rangle_n^c 
- \beta_o\left\langle \tilde{t}\tilde{x}\right\rangle_{n-1}^c 
\nonumber \\ &&
- \beta_o\left\langle \tilde{t}\tilde{x}\right\rangle_{n+1}^c 
- \beta_o\left\langle \tilde{t}\tilde{y}\right\rangle_{n+1}^s 
+\beta_o\left\langle \tilde{t}\tilde{y}\right\rangle_{n-1}^s
\\
(R_o^2)_n^s & = & 
\frac{1}{2}\left\langle \tilde{x}^2\right\rangle _n^s
+\frac{1}{2}\left\langle \tilde{y}^2\right\rangle _n^s
-\frac{1}{4}\left\langle \tilde{y}^2\right\rangle _{n-2}^s
+\frac{1}{4}\left\langle \tilde{y}^2\right\rangle _{n+2}^s
\nonumber \\ && 
+\frac{1}{4}\left\langle \tilde{x}^2\right\rangle _{n-2}^s
-\frac{1}{4}\left\langle \tilde{x}^2\right\rangle _{n+2}^s
+\frac{1}{2}\left\langle \tilde{x}\tilde{y}\right\rangle _{n-2}^c
\nonumber \\ &&
-\frac{1}{2}\left\langle \tilde{x}\tilde{y}\right\rangle _{n+2}^c 
+\beta_o^2\left\langle \tilde{t}^2\right\rangle_n^s 
- \beta_o\left\langle \tilde{t}\tilde{x}\right\rangle_{n-1}^s 
\nonumber \\ &&
- \beta_o\left\langle \tilde{t}\tilde{x}\right\rangle_{n+1}^s 
- \beta_o\left\langle \tilde{t}\tilde{y}\right\rangle_{n-1}^c 
+\beta_o\left\langle \tilde{t}\tilde{y}\right\rangle_{n+1}^c \, .
\end{eqnarray}
\end{subequations}
Note that from the third order onwards we have given general expressions for any order. 

Analogically we derived the series for the sideward radius.
\allowdisplaybreaks[1]
\begin{subequations}
\begin{eqnarray}
(R_s^2)_0 & = & 
\frac{1}{2}\left\langle \tilde{x}^2\right\rangle _0
+\frac{1}{2}\left\langle \tilde{y}^2\right\rangle _0
+\frac{1}{4}\left\langle \tilde{y}^2\right\rangle _2^c
-\frac{1}{4}\left\langle \tilde{x}^2\right\rangle _2^c
\nonumber \\ && 
-\frac{1}{2}\left\langle \tilde{x}\tilde{y}\right\rangle _2^s 
\\
(R_s^2)_1^c & = & 
\frac{1}{4}\left\langle \tilde{x}^2\right\rangle _1^c
+\frac{3}{4}\left\langle \tilde{y}^2\right\rangle _1^c
+\frac{1}{4}\left\langle \tilde{y}^2\right\rangle _3^c
-\frac{1}{4}\left\langle \tilde{x}^2\right\rangle _3^c
\nonumber \\ &&
-\frac{1}{2}\left(\left\langle \tilde{x}\tilde{y}\right\rangle _1^s
-\left\langle \tilde{x}\tilde{y}\right\rangle _3^s\right) 
\\
(R_s^2)_1^s & = & 
\frac{3}{4}\left\langle \tilde{x}^2\right\rangle _1^s
+\frac{1}{4}\left\langle \tilde{y}^2\right\rangle _1^s
+\frac{1}{4}\left\langle \tilde{y}^2\right\rangle _3^s
-\frac{1}{4}\left\langle \tilde{x}^2\right\rangle _3^s
\nonumber \\ && 
-\frac{1}{2}\left(\left\langle \tilde{x}\tilde{y}\right\rangle _1^c
-\left\langle \tilde{x}\tilde{y}\right\rangle _3^c\right) 
\\
(R_s^2)_2^c & = & 
\frac{1}{2}\left\langle \tilde{x}^2\right\rangle _2^c
+\frac{1}{2}\left\langle \tilde{y}^2\right\rangle _2^c
+\frac{1}{2}\left\langle \tilde{y}^2\right\rangle _0
+\frac{1}{4}\left\langle \tilde{y}^2\right\rangle _4^c
\nonumber \\ && 
-\frac{1}{2}\left\langle \tilde{x}^2\right\rangle _0
-\frac{1}{4}\left\langle \tilde{x}^2\right\rangle_4^c
-\frac{1}{2}\left\langle \tilde{x}\tilde{y}\right\rangle _4^s 
\\
(R_s^2)_2^s & = & 
\frac{1}{2}\left\langle \tilde{x}^2\right\rangle _2^s
+\frac{1}{2}\left\langle \tilde{y}^2\right\rangle _2^s
+\frac{1}{4}\left\langle \tilde{y}^2\right\rangle _4^s
-\frac{1}{4}\left\langle \tilde{x}^2\right\rangle_4^s
\nonumber \\ && 
-\left\langle \tilde{x}\tilde{y}\right\rangle _0
+\frac{1}{2}\left\langle \tilde{x}\tilde{y}\right\rangle _4^c 
\\
(R_s^2)_n^c & = & 
\frac{1}{2}\left\langle \tilde{x}^2\right\rangle _n^c
+\frac{1}{2}\left\langle \tilde{y}^2\right\rangle _n^c
+\frac{1}{4}\left\langle \tilde{y}^2\right\rangle _{n-2}^c
+\frac{1}{4}\left\langle \tilde{y}^2\right\rangle _{n+2}^c
\nonumber \\ && 
-\frac{1}{4}\left\langle \tilde{x}^2\right\rangle _{n-2}^c
-\frac{1}{4}\left\langle \tilde{x}^2\right\rangle _{n+2}^c 
-\frac{1}{2}\left\langle \tilde{x}\tilde{y}\right\rangle _{n+2}^s
\nonumber \\ && 
+\frac{1}{2}\left\langle \tilde{x}\tilde{y}\right\rangle _{n-2}^s 
\\
(R_s^2)_n^s & = & 
\frac{1}{2}\left\langle \tilde{x}^2\right\rangle _n^s
+\frac{1}{2}\left\langle \tilde{y}^2\right\rangle _n^s
+\frac{1}{4}\left\langle \tilde{y}^2\right\rangle _{n-2}^s
+\frac{1}{4}\left\langle \tilde{y}^2\right\rangle _{n+2}^s
\nonumber \\ && 
-\frac{1}{4}\left\langle \tilde{x}^2\right\rangle _{n-2}^s
-\frac{1}{4}\left\langle \tilde{x}^2\right\rangle _{n+2}^s 
-\frac{1}{2}\left\langle \tilde{x}\tilde{y}\right\rangle _{n-2}^c
\nonumber \\ && 
+\frac{1}{2}\left\langle \tilde{x}\tilde{y}\right\rangle _{n+2}^c \, .
\end{eqnarray}
\end{subequations}

Finally, for the cross-term we obtained
\begin{subequations}
\begin{eqnarray}
(R_{os}^2)_0 & = & 
\frac{1}{2}\left\langle \tilde{x}\tilde{y}\right\rangle _2^c
+\frac{1}{4}\left\langle \tilde{y}^2\right\rangle _2^s
-\frac{1}{4}\left\langle \tilde{x}^2\right\rangle _2^s
+\frac{\beta_o}{2}\left\langle \tilde{x}\tilde{t}\right\rangle _1^s
\nonumber \\ &&
-\frac{\beta_o}{2}\left\langle \tilde{y}\tilde{t}\right\rangle _1^c 
\\
(R_{os}^2)_1^c & = & 
\frac{1}{2}\left\langle \tilde{x}\tilde{y}\right\rangle _1^c
+\frac{1}{2}\left\langle \tilde{x}\tilde{y}\right\rangle _3^c
+\frac{1}{4}\left\langle \tilde{y}^2\right\rangle _1^s
+\frac{1}{4}\left\langle \tilde{y}^2\right\rangle _3^s
\nonumber \\ &&
-\frac{1}{4}\left\langle \tilde{x}^2\right\rangle _1^s
-\frac{1}{4}\left\langle \tilde{x}^2\right\rangle _3^s 
+ \frac{\beta_o}{2}\left\langle \tilde{x}\tilde{t}\right\rangle_2^s
- \beta_o\left\langle \tilde{y}\tilde{t}\right\rangle_0
\nonumber \\ &&
 - \frac{\beta_o}{2}\left\langle \tilde{y}\tilde{t}\right\rangle_2^c 
 \\
(R_{os}^2)_1^s & = & 
-\frac{1}{2}\left\langle \tilde{x}\tilde{y}\right\rangle _1^s
-\frac{1}{2}\left\langle \tilde{x}\tilde{y}\right\rangle _3^s
+\frac{1}{4}\left\langle \tilde{y}^2\right\rangle _1^c
-\frac{1}{4}\left\langle \tilde{y}^2\right\rangle _3^c
\nonumber \\ &&
-\frac{1}{4}\left\langle \tilde{x}^2\right\rangle _1^c
+\frac{1}{4}\left\langle \tilde{x}^2\right\rangle _3^c 
+\beta_o\left\langle \tilde{x}\tilde{t}\right\rangle_0 
- \frac{\beta_o}{2}\left\langle \tilde{x}\tilde{t}\right\rangle_2^c 
\nonumber \\ &&
- \frac{\beta_o}{2}\left\langle \tilde{y}\tilde{t}\right\rangle_2^s
\\
(R_{os}^2)_2^c & = & 
\left\langle \tilde{x}\tilde{y}\right\rangle _0
+\frac{1}{2}\left\langle \tilde{x}\tilde{y}\right\rangle _4^c
+\frac{1}{4}\left\langle \tilde{y}^2\right\rangle _4^s
-\frac{1}{4}\left\langle \tilde{x}^2\right\rangle _4^s 
\nonumber \\ &&
- \frac{\beta_o}{2}\left\langle \tilde{x}\tilde{t}\right\rangle_1^s
+ \frac{\beta_o}{2}\left\langle \tilde{x}\tilde{t}\right\rangle_3^s 
- \frac{\beta_o}{2}\left\langle \tilde{y}\tilde{t}\right\rangle_1^c 
\nonumber \\ &&
- \frac{\beta_o}{2}\left\langle \tilde{y}\tilde{t}\right\rangle_3^c 
\\
(R_{os}^2)_2^s & = & 
-\frac{1}{2}\left\langle \tilde{x}\tilde{y}\right\rangle _4^s
+\frac{1}{2}\left\langle \tilde{y}^2\right\rangle _0
-\frac{1}{4}\left\langle \tilde{y}^2\right\rangle _4^c
-\frac{1}{2}\left\langle \tilde{x}^2\right\rangle _0
\nonumber \\ &&
+\frac{1}{4}\left\langle \tilde{x}^2\right\rangle _4^c 
+\frac{\beta_o}{2}\left\langle \tilde{x}\tilde{t}\right\rangle_1^c 
- \frac{\beta_o}{2}\left\langle \tilde{x}\tilde{t}\right\rangle_3^c 
- \frac{\beta_o}{2}\left\langle \tilde{y}\tilde{t}\right\rangle_1^s
\nonumber \\ &&
- \frac{\beta_o}{2}\left\langle \tilde{y}\tilde{t}\right\rangle_3^s
\\
(R_{os}^2)_n^c & = & 
\frac{1}{2}\left\langle \tilde{x}\tilde{y}\right\rangle _{n-2}^c
+\frac{1}{2}\left\langle \tilde{x}\tilde{y}\right\rangle _{n+2}^c
+\frac{1}{4}\left\langle \tilde{y}^2\right\rangle _{n+2}^s
\nonumber \\ &&
-\frac{1}{4}\left\langle \tilde{y}^2\right\rangle _{n-2}^s
-\frac{1}{4}\left\langle \tilde{x}^2\right\rangle _{n+2}^s
+\frac{1}{4}\left\langle \tilde{x}^2\right\rangle _{n-2}^s
\nonumber \\ && 
+ \frac{\beta_o}{2}\left\langle \tilde{x}\tilde{t}\right\rangle_{n+1}^s
- \frac{\beta_o}{2}\left\langle \tilde{x}\tilde{t}\right\rangle_{n-1}^s  
- \frac{\beta_o}{2}\left\langle \tilde{y}\tilde{t}\right\rangle_{n+1}^c 
\nonumber \\ &&
- \frac{\beta_o}{2}\left\langle \tilde{y}\tilde{t}\right\rangle_{n-1}^c 
\\
(R_{os}^2)_n^s & = & 
-\frac{1}{2}\left\langle \tilde{x}\tilde{y}\right\rangle _{n+2}^s
+\frac{1}{2}\left\langle \tilde{x}\tilde{y}\right\rangle _{n-2}^s
-\frac{1}{4}\left\langle \tilde{y}^2\right\rangle _{n+2}^c
\nonumber \\ &&
+\frac{1}{4}\left\langle \tilde{y}^2\right\rangle _{n-2}^c
+\frac{1}{4}\left\langle \tilde{x}^2\right\rangle _{n+2}^c
-\frac{1}{4}\left\langle \tilde{x}^2\right\rangle _{n-2}^c 
\nonumber \\ &&
- \frac{\beta_o}{2}\left\langle \tilde{x}\tilde{t}\right\rangle_{n+1}^c
+ \frac{\beta_o}{2}\left\langle \tilde{x}\tilde{t}\right\rangle_{n-1}^c  
- \frac{\beta_o}{2}\left\langle \tilde{y}\tilde{t}\right\rangle_{n+1}^s
\nonumber \\ && 
- \frac{\beta_o}{2}\left\langle \tilde{y}\tilde{t}\right\rangle_{n-1}^s 
\,  .
\end{eqnarray}
\end{subequations}

%%%%%%%%%%%%%%%%%%%%%%%%%%%%%%%%%%%%%%%%%%%%%%%%%%

\end{document}